\documentstyle{mn}
\def\spose#1{\hbox to 0pt{#1\hss}}
\def\simlt{\mathrel{\spose{\lower 3pt\hbox{$\mathchar"218$}}
     \raise 2.0pt\hbox{$\mathchar"13C$}}}
\def\simgt{\mathrel{\spose{\lower 3pt\hbox{$\mathchar"218$}}
     \raise 2.0pt\hbox{$\mathchar"13E$}}}
\def\HST{{\it HST\/}}
\newcommand{\II}{{\footnotesize~II}}
\newcommand{\III}{{\footnotesize~III}}

\begin{document}
\title{Thermal-infrared imaging of 3C radio galaxies at $z\sim1$}
\author[C. Simpson et al.]{Chris Simpson,$^1$ Steve Rawlings$^2$ and Mark
Lacy$^2$\\
$^1$Subaru Telescope, National Astronomical Observatory of Japan, 650
N.~A`Oh\={o}k\={u} Place, Hilo, HI 96720, U.S.A.\\
$^2$Astrophysics, Department of Physics University of Oxford, Keble Road,
Oxford OX1 3RH}
\maketitle

\begin{abstract}\normalsize
We present the results of a programme of thermal-infrared imaging of
nineteen $z \sim 1$ radio galaxies from the 3CR and 3CRR (LRL)
samples. We detect emission at $L'$ ($3.8\,\mu$m) from four objects;
in each case the emission is unresolved at 1-arcsec resolution.
Fifteen radio galaxies remain undetected to sensitive limits of $L'
\approx 15.5$. Using these data in tandem with archived \HST\ data and
near-infrared spectroscopy we show that three of the detected `radio
galaxies' (3C~22, 3C~41, and 3C~65) harbour quasars reddened by $A_V
\simlt 5$. Correcting for this reddening 3C~22 and 3C~41 are very
similar to coeval 3C quasars, whilst 3C~65 seems unusually
underluminous. The fourth radio galaxy detection (3C~265) is a more
highly obscured ($A_V \sim 15$) but otherwise typical quasar which
previously has been evident only in scattered light. We determine the
fraction of dust-reddened quasars at $z \sim 1$ to be
$28^{+25}_{-13}$\% at 90\% confidence. On the assumption that the
undetected radio galaxies harbour quasars similar to those in 3C~22,
3C~41 and 3C~265 (as seems reasonable given their similar narrow
emission line luminosities) we deduce extinctions of $A_V \simgt 15$
towards their nuclei. The contributions of reddened quasar nuclei to
the total $K$-band light ranges from $\sim 0$ per cent for the
non-detections, through $\sim 10$ per cent for 3C~265 to $\sim 80$ per
cent for 3C~22 and 3C~41. Correcting for these effects does not remove
the previously reported differences between the $K$ magnitudes of 3C
and 6C radio galaxies, so contamination by reddened quasar nuclei is
not a serious problem for drawing cosmological conclusions from the
$K$--$z$ relation for radio galaxies. We discuss these results in the
context of the `receding torus' model which predicts a small fraction
of lightly-reddened quasars in samples of high radio luminosity
sources. We also examine the likely future importance of
thermal-infrared imaging in the study of distant powerful radio
sources.

\end{abstract}

\section{Introduction}
\label{sec:intro}

Luminous extragalactic radio sources are associated with either
quasars or radio galaxies, the latter class of object lacking the
bright non-stellar continuum and prominent broad emission lines that
characterize the former. For many years astronomers have speculated
that at least some radio galaxies harbour an active nucleus which,
although obscured from view along the line of sight, would be observed
as a quasar if seen from certain other, favourable directions (e.g.,
Scheuer 1987). There is now overwhelming observational evidence that
this is the case. Most notably, spectropolarimetry has revealed the
presence of quasar-like broad emission lines in the polarized flux
spectra of a number of radio galaxies (e.g., Tran et al.\ 1998 and
references therein). In such cases the broad lines are hidden from
direct view but are scattered into our line of sight by a screen of
either dust or electrons. However, the more general proposition that
{\em all\/} radio galaxies harbour obscured quasar nuclei has not yet
been confirmed.

It is generally believed that the unobscured lines of sight lie close
to the axis of the twin radio jets, and that quasars are therefore
seen more nearly `pole-on' than their radio galaxy counterparts (e.g.,
Barthel 1989). This is supported by the presence of bright radio cores
and one-sided jets in quasars that are weak or absent in radio
galaxies (e.g., Owen \& Puschell 1984), a result which is most
naturally explained by Doppler boosting of emission from relativistic
jets moving close to the line of sight. Even more compelling evidence
comes from optical polarization studies that have revealed many radio
galaxies in which the rest-frame ultraviolet light is strongly
polarized with its electric vector lying perpendicular to the radio
axis (e.g., Tadhunter et al.\ 1992). These observations suggest that
quasar light can emerge relatively unhindered along the radio axis,
and reach the observer after scattering, whilst direct transmitted
radiation from the quasar nucleus is effectively blocked. Such
processes are also at least partly responsible for the so-called
`alignment effect' (Chambers, Miley \& van Breugel 1987; McCarthy et
al.\ 1987) --- the tendency for the elongation of optical and radio
structures along a common axis.

In an orientation-based unification scheme such as this, the obscuring
material lies preferentially in the plane perpendicular to the radio
jets, and a central parameter is the angle between the radio axis and
the line of sight which grazes the edge of the obscuring material
(generally referred to as the `torus').  A simple model which is often
considered relates this critical angle $\theta_{\rm c}$ to the inner
radius of the torus, $r$, and its half-height, $h$, via $\tan
\theta_{\rm c} = r/h$ (see, e.g., Simpson 1998). In this model there
are physical reasons why $r$, and hence $\theta_{\rm c}$, might be
expected to correlate with quasar luminosity: if the value of $r$ is
determined by the radius at which dust sublimates in the radiation
field of the quasar then it should increase with the photoionizing
luminosity of the quasar as $L_{\rm phot}^{0.5}$ (Lawrence 1991).
This picture where the torus opens up as the luminosity of the quasar
increases in known as the `receding torus' model. Strong links between
the intrinsic UV/optical luminosity of the quasar nuclei of radio
sources and extended radio luminosity are now firmly established
(Rawlings \& Saunders 1991; Falcke, Malkan \& Biermann 1995; Serjeant
et al.\ 1998; Willott et al.\ 1998c) in the sense that $L_{\rm phot}
\propto L_{178}^{0.6}$ where $L_{178}$ is the 178-MHz (extended) radio
luminosity.  It therefore seems highly plausible that $r$ and hence
$\theta_{\rm c}$ will rise, albeit slowly, with radio luminosity
($\tan \theta_{\rm c} \propto L_{178}^{0.3}$),

The brightest (3C) radio sources at $z \sim 1$ are about 100 times
more radio luminous than 3C sources at $z \sim 0.1$, so assuming $h$
to be independent of $L_{178}$ and $z$, we expect $r$ and hence $\tan
\theta_{\rm c}$ to be a factor $\sim 4$ greater for these objects. The
quasar fraction should therefore be correspondingly larger, and, since
lines of sight grazing the torus will subtend a much lower solid
angle, the fraction of lightly-reddened quasars should be much reduced
(see Hill et al.\ 1996). Based on their detection of a high fraction
of lightly-reddened quasar nuclei from near-infrared spectroscopy of
Pa$\alpha$ in low-redshift ($0.1 \leq z < 0.2$) 3C radio sources,
Hill, Goodrich \& DePoy (1996) found $\theta_{\rm c} \approx
20^\circ$, and used the receding torus model to make quantitative
predictions about the distribution of nuclear extinctions in
higher-redshift, more radio luminous, sources. Their predictions were
in rough agreement with Economou et al.'s (1995) discovery of only one
broad H$\alpha$ line in a preliminary analysis of a sample of ten
radio galaxies at $z \sim 1$. However, as Rawlings et al.\ (1995)
discuss, spectroscopic methods of detecting hidden quasar nuclei lack
the sensitivity to provide a good general method of characterizing the
angular distribution of reddening towards quasar nuclei in radio
galaxies at $z \sim 1$. This is because of the large optical depth
expected towards rest-frame optical lines like H$\alpha$, and the poor
sensitivity to weak rest-frame near-infrared lines like Pa$\alpha$,
particularly once they become redshifted beyond the $K$-band (at $z >
0.3$ in the case of Pa$\alpha$). A definitive test of the receding
torus model for high-redshift objects has therefore yet to be
performed.

\begin{figure}
\vspace*{125mm}
\includegraphics{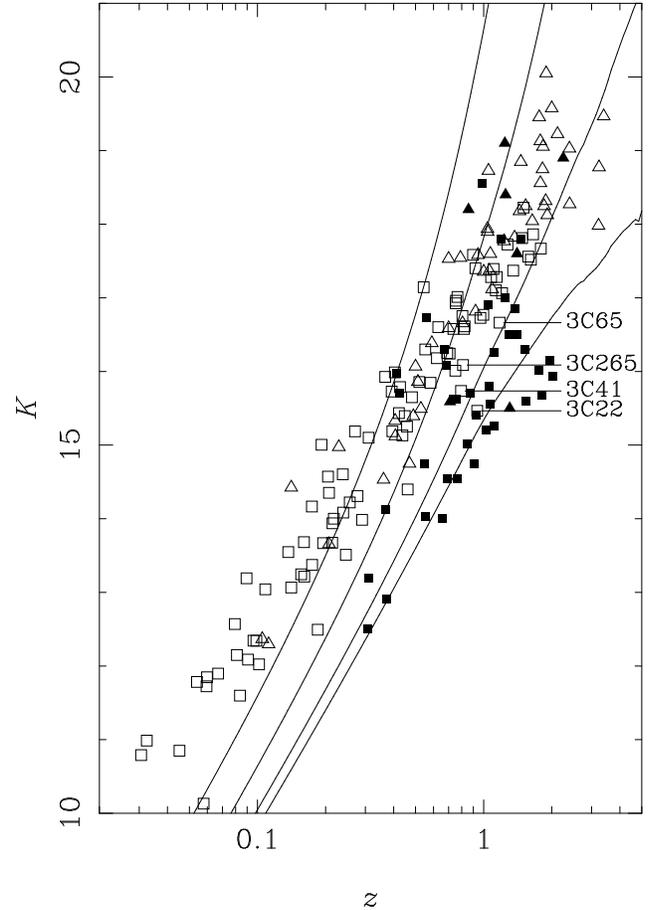}
\caption[]{The near-infrared Hubble Diagram for radio sources. The
objects are taken from the 3CRR (LRL) sample (squares; Laing, Riley \&
Longair 1983), and the 6C sample (triangles; Eales 1985, slightly
revised by Rawlings et al.\ 1998b); quasars are represented by filled
symbols, and other objects by open symbols. We define quasars are
having a nuclear point source with $M_B < -23$; with this criterion,
all quasars display broad optical emission lines, but some of the
other objects, which we term radio galaxies, also display broad lines,
either weakly or in polarized flux only. $K$-band photometry for the
LRL sample has been taken from Best et al.\ (1998), Lilly \& Longair
(1982, 1984), Lilly, Longair \& Miller (1985) and Rawlings et al.\
(1995).  Although 73 of the 94 $z > 0.1$ LRL radio galaxies have
$K$-band photometry (and in all but one case the lack of $K$-band data
stems from the inability to observe high-declination objects at
UKIRT), almost all the quasars lack such data; to plot the quasars we
have estimated $K$-band magnitudes using optical photometry from LRL,
and an extrapolation based on the canonical quasar spectrum described
in Section~\ref{sec:intro}. Note that contributions from the host
galaxies of quasars, and the possibility of their suffering
significant reddening mean that these estimated $K$ values should be
treated as upper limits. $K$-band photometry and redshifts for 6C
sources are from Eales et al.\ (1997) and Rawlings et al.\ (1998b)
respectively.  39 of the 41 $z > 0.1$ 6C radio galaxies in the 6C
sample have $K$-band photometry; we again estimate $K$-band magnitudes
for the 7 quasars using optical photometry (Eales 1985) and the
canonical quasar spectrum. The lines show the loci of a model quasar
with $M_B=-26$ reddened by $A_V=0,2,7,15$; the lowest curve is for
$A_V=0$, increasingly higher curves represent increasing values of
$A_V$. A few objects of special interest are labelled.}
\label{fig:kz}
\end{figure}

The distribution of reddenings to the quasar nuclei may also have
important cosmological implications, since the near-infrared Hubble
diagram for radio galaxies (Fig.~\ref{fig:kz}) --- and the common
assumption that the near-infrared emission of radio galaxies is
dominated by starlight --- has been used to infer constraints on the
formation and evolution of elliptical galaxies (e.g., Lilly 1989;
Eales et al.\ 1993). The discovery that the $K$-band magnitudes of 6C
radio galaxies, i.e., those selected at a $\sim 6$ times lower radio
flux limit than 3C, are $\sim 0.6$ magnitudes fainter than those of 3C
radio galaxies at the same redshift (Eales et al.\ 1997) suggests that
there is some close link between radio luminosity and near-infrared
luminosity.  Either, stellar luminosity --- and by inference stellar
mass --- correlates with radio luminosity, or there is some important
component of the near-infrared luminosity which can be ascribed to a
hidden quasar nucleus. Recent discussions of these possibilities can
be found in Eales et al.\ (1997), Best, Longair \& R\"{o}ttgering
(1998) and Roche, Eales \& Rawlings (1998).  Although there is direct
observational evidence from near-infrared polarimetry (Leyshon \&
Eales 1998) that scattered light makes only a small (few per cent)
contribution to the near-infrared luminosity of radio galaxies even
when it makes a very significant contribution in the ultraviolet and
optical, it has not yet been conclusively shown that reddened quasar
light is also a negligible component. As illustrated by
Fig.~\ref{fig:kz}, such quasar nuclei would need to be subject to only
a fairly limited amount of reddening ($A_V \sim 2$--7) in order not to
possess obvious quasar features in their optical spectra whilst still
contributing to the $K$-band luminosity at the level of radio galaxies
lying close to the mean $K$--$z$ relation for radio galaxies. The
existence of a substantial population of moderately-reddened quasars
has been suggested by Webster et al.\ (1995), although both the
reality and importance of this population have been questioned
(Serjeant \& Rawlings 1996; Benn et al.\ 1998).

\begin{figure}
\vspace*{126mm}
\includegraphics{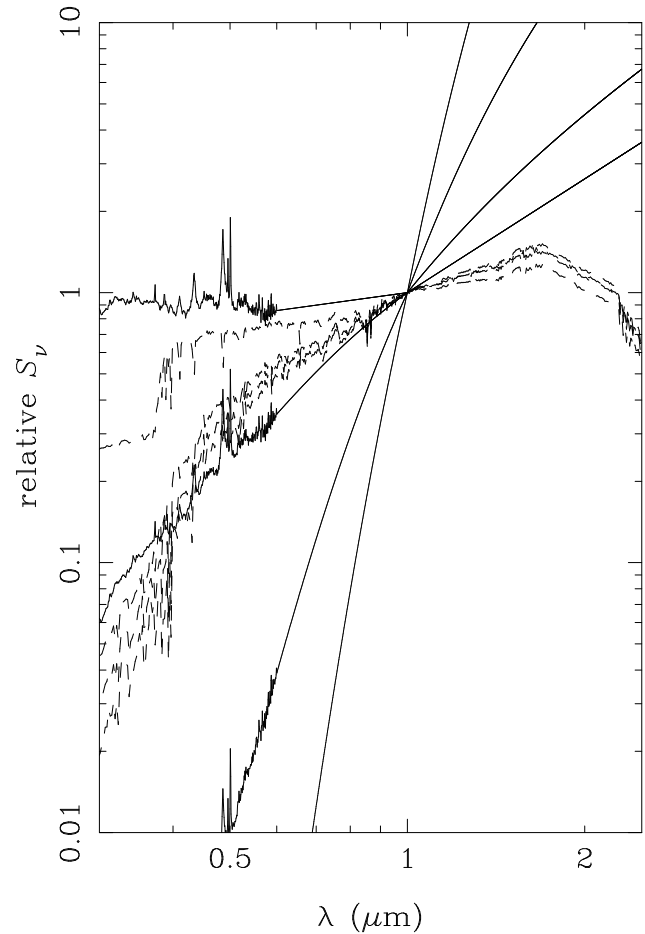}
\caption[]{Comparison of the spectral energy
distributions SEDs of stellar populations (of various ages) and
quasars (subject to various amounts of reddening). Stellar population
loci (dashed curves) are 1-Gyr bursts at ages 1,2,3, and 5\,Gyr
(younger ages correspond to bluer colours) from the GISSEL96 models
(Bruzual \& Charlot 1993,1998).  Quasar loci (full lines) have
been calculated using the canonical spectrum described in Section~1 as
seen through 0,2,7, and 15\,mag of visual extinction in the rest-frame
of the quasar host galaxy (lower extinctions correspond to bluer
colours).  All the SEDs have been normalized at 1\,$\mu$m. Note the
similarities between the SEDs of lightly-reddened quasars ($A_V \sim
2$) and old stellar populations blueward of 1\,$\mu$m, but the
divergence at longer wavelengths.}
\label{fig:seds}
\end{figure}

In this paper we report on the results from a project designed with
two principal aims. The first aim was to determine the contribution of
transmitted quasar light to the $K$--$z$ relation for $z \sim 1$ radio
galaxies, and hence determined whether reddened quasar nuclei can
explain the difference in near-infrared luminosities between 3C and 6C
radio galaxies. The second aim was to constrain the range of $A_V$
towards the quasar nuclei of luminous radio sources at $z \sim 1$, and
to compare this range with the predictions of the receding torus
model.  The project was based around thermal-infrared imaging since,
as discussed by Rawlings et al.\ (1995) this is for these purposes a
much more powerful technique than spectroscopy. Its utility rests on
the dramatic divergence of the rest-frame spectral energy
distributions (SEDs) of starlight and quasar light redward of
1\,$\mu$m (see Fig.~\ref{fig:seds}).

Although there is considerable variation in the quantitative spectral
energy distributions of quasars, the inflexion at 1\,$\mu$m appears to
be a universal feature (Neugebauer et al.\ 1987; Sanders et al.\ 1989;
Elvis et al.\ 1994), except for blazar-type sources where synchrotron
emission dominates even at optical--infrared wavelengths. The rise to
longer wavelengths is believed to be due to dust at or near its
sublimation temperature ($T \sim 1500$\,K). Kobayashi et al.\ (1993)
have successfully modelled the near-infrared continua of low-redshift
quasars in this manner, and Barvainis (1987) has shown that the entire
infrared spectrum from 1--100\,$\mu$m can be attributed to dust at a
range of temperatures, with the hottest dust contributing most
strongly at the shortest wavelengths. If the dust is heated by, and in
thermal equilibrium with, the radiation field of the quasar, then the
hottest dust must originate close to the central engine. Thus even
though emission from the hot dust may be optically thin and hence
isotropic, it could easily be occluded by the larger-scale obscuring
material so that the emergent flux becomes a strong function of
viewing angle. The attenuation suffered at angles greater than
$\theta_{\rm c}$ should, however, be much less severe than at shorter
wavelengths.  The apparent constancy of the near-infrared continuum
slope in quasars (Neugebauer et al. 1987), and the possibility that
the emitting regions are optically thin makes it an ideal region of
the spectrum from which to determine the reddening.

The structure of this paper is as follows.  In Section~2 we describe
our selection of targets for thermal-infrared imaging.  In Section~3
we describe our new imaging observations with the 3.9-m United Kingdom
Infrared Telescope (UKIRT). In Section~4 we present the results of
these observations and our analysis of the data. In Section~5 we widen
our analysis to include archival data from the Hubble Space Telescope
(\HST), as well as a new near-infrared spectrum obtained at UKIRT. In
Section~6 we discuss the implications of our results, including both a
discussion of the effects of hidden quasar nuclei on the $K$--$z$
relation and a test of the receding torus model. In Section~7 we
review the prospects for future observations of the type discussed in
this paper.

Throughout this paper we have assumed $H_0 =
50$\,km\,s$^{-1}$\,Mpc$^{-1}$, $q_0 = 0.5$ and $\Lambda = 0$. The
convention for spectral index, $\alpha$, is that $S_\nu \propto
\nu^{-\alpha}$, where $S_\nu$ is the flux density at frequency
$\nu$. We have adopted a canonical rest-frame spectrum for a quasar of
the following form: the composite quasar spectrum from Francis et al.\
(1991) for wavelengths from 0.1--0.6\,$\mu$m, a power law with $\alpha
= 0.3$ from 0.6--1\,$\mu$m, and a power law with $\alpha = 1.4$ from
1--3\,$\mu$m. The motivation for the broken power-law form at near-IR
wavelengths is provided by the data of Neugebauer et al.\ (1987). We
adopt the empirical extinction curve for the Milky Way given by Pei
(1992) as our assumed dust extinction curve.

\section{Selection of targets}

Our primary goal in this project was to study a complete subsample of
radio galaxies from the 3CRR sample --- which we hereafter refer to as
the LRL sample (after Laing, Riley \& Longair 1983) in order to more
easily differentiate it from the 3CR sample --- in the redshift range
$0.65 \leq z < 1.20$.  These objects have 178-MHz radio luminosities
$L_{178} > 10^{27}$\,W\,Hz$^{-1}$\,sr$^{-1}$, and they lie at
redshifts where 3.8\,$\mu$m imaging probes rest-frame wavelengths in
the region of primary interest (see Fig.~\ref{fig:seds}). However, a
combination of scheduling, technical problems, and poor weather meant
that we were often either unable to observe all the sources within a
given patch of sky, or conversely were short of targets. In the former
case, we chose the targets to observe based solely on position in the
sky so as to make our observing programme as efficient as possible,
while in the latter case, we supplemented our list from the 3CR
catalogue of Bennett (1962a,b), on which LRL is based. The declination
limit of UKIRT forced us to exclude targets with $\delta >
+60^\circ$. Over the course of the three observing runs assigned to
this project, plus some additional spare time on three other nights,
we were able to observe all twelve radio galaxies from 3CR and LRL
satisfying $\alpha > 16^{\rm h}$ and $\alpha < 04^{\rm h} 20^{\rm
m}$. Our coverage in the RA range $09^{\rm h} < \alpha < 14^{\rm h}$
is less complete. We observed a randomly-selected subsample of five 3C
radio galaxies from the eight with $0.8 < z < 1.0$; the excluded
sources were 3C~237 (in 3CR only) and 3C~263.1 and 3C~280 (also in
LRL). We provide relevant data for our 3CR (but non-LRL) targets in
Table~\ref{tab:nonlrl}; a useful recent reference for data on the LRL
sample is Blundell et al.\ (1998). A full list of our targets is
included in Table~\ref{tab:log}

\begin{table*}
\centering
\caption[]{Relevant data on the UKIRT 3CR targets which are not in the
LRL sample. Positions are either optical (from McCarthy et al.\
1997) or radio (from Strom et al.\ 1990). We provide radio core
measurments (and the frequency at which these measurements were made)
and the largest (projected) angular sizes of the radio sources from
the following references: B94, Bogers et al.\ (1994); L75, Longair
(1975); R96, Rhee et al.\ (1996); S90, Strom et al.\ (1990).}
\label{tab:nonlrl}
\begin{tabular}{llllrccl}
Name & Pos &
\multicolumn{2}{c}{$\alpha$\hspace{10pt}(B1950)\hspace{10pt}$\delta$}
&  \multicolumn{1}{c}{$S_{\rm core}$ (mJy)} &
\multicolumn{1}{c}{$\nu$ (GHz)} & LAS (\arcsec) &
\multicolumn{1}{c}{Ref} \\
\hline
3C~44  & O$^*$ & 01 28 45.12 & $+$06 08 17.7 & $<1.0$ & 8.0 & 65 & B94 \\
3C~54  & O     & 01 52 25.91 & $+$43 31 20.9 &$<25.0$ & 5.0 & 54 & L75 \\
3C~107 & O     & 04 09 49.88 & $-$01 07 10.6 &$<13.0$ & 1.4 & 13 & R96 \\
3C~114 & R     & 04 17 29.05 & $+$17 46 49.0 &  16.0  & 5.0 & 54 & S90 \\
3C~272 & R     & 12 22 00.81 & $+$42 23 12.3 &   0.3  & 5.0 & 59 & S90 \\
\hline
\end{tabular}

$^*$Position estimated from the countour plot in McCarthy et al.\
(1997) since the position given in their Table~3 is incorrect.
\end{table*}

\begin{figure}
\vspace*{128mm}
\includegraphics{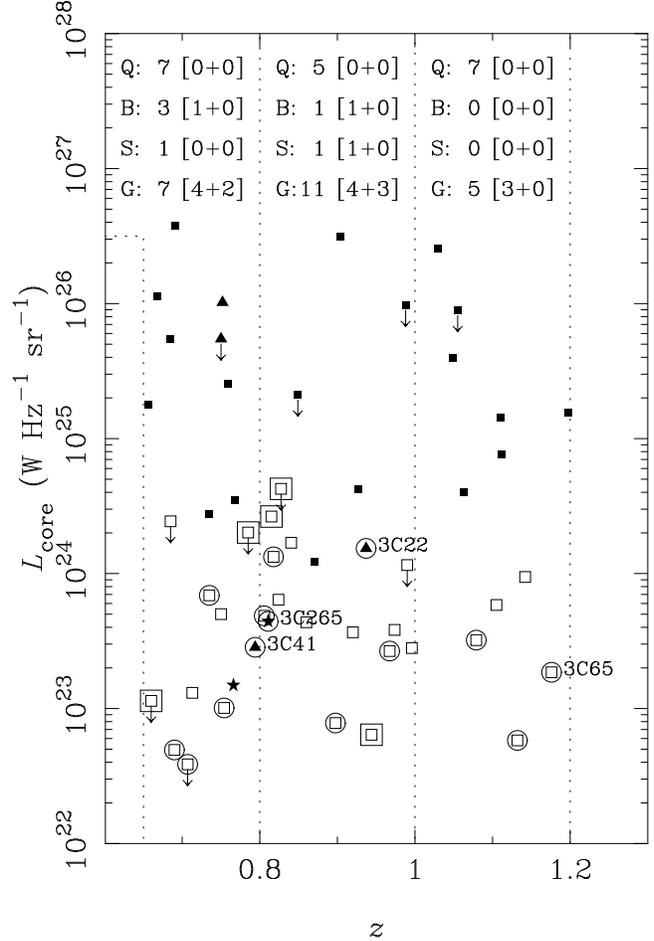}
\caption[]{Radio core luminosity (evaluated at or
near 5\,GHz) versus redshift for all objects from the LRL sample with
$0.65 \leq z < 1.20$ plus additional UKIRT targets.  UKIRT targets are
indicated by large symbols: large circles denoting LRL objects; and
large squares, objects which are only in 3CR. The smaller symbols have
meanings as follows: filled squares represent quasars (class `Q');
filled triangles, radio galaxies with broad lines prominent in
direct-light spectra (`B'); filled stars, radio galaxies with broad
lines seen only because of a large scattered (polarized) component
(`S'); and open squares, narrow-line radio galaxies (`G').  The
numbers in each category in each of three marked redshift bins are
summarized at the top of the plot, with the first number relating to
the total number in LRL, the second to the number of LRL objects
observed with the UKIRT, and the third number to the 3CR (only) UKIRT
targets.  A few objects of special interest are labelled.}
\label{fig:cores}
\end{figure}

In Fig.~\ref{fig:cores}, we compare the location of our UKIRT targets
with the full LRL sample in a plot of radio core luminosity versus
redshift.  This figure introduces two new classes containing objects
traditionally classified as radio galaxies (e.g., in LRL) but with
broad emission lines in their optical and/or near-infrared
spectra. The first new class accounts for objects with prominent broad
lines in their direct-light spectra but nuclear optical continuum
luminosities below the $M_{B} = -23$ limit often taken as the minimum
required for a quasar (hereafter class-`B' radio galaxies; see Willott
et al. 1998b for further discussion).  Since these criteria ignore any
reddening corrections, class-`B' radio galaxies include both
lightly-reddened quasars and those that are simply intrinsically weak.
The second class accounts for objects with broad emission lines which
are prominent only in scattered light as evidenced by
spectropolarimetric observations (hereafter class-`S' radio galaxies;
e.g., Tran et al.\ 1998).  Class-`S' objects obviously harbour quasar
nuclei, but according to unified schemes may be being observed very
far from a pole-on orientation.  Fig.~\ref{fig:cores} illustrates the
dichotomy between the radio core properties of quasars and narrow-line
radio galaxies and hints that class-`B' (and maybe class-`S') radio
galaxies are intermediary cases.

\begin{figure}
\vspace*{410pt}
\includegraphics{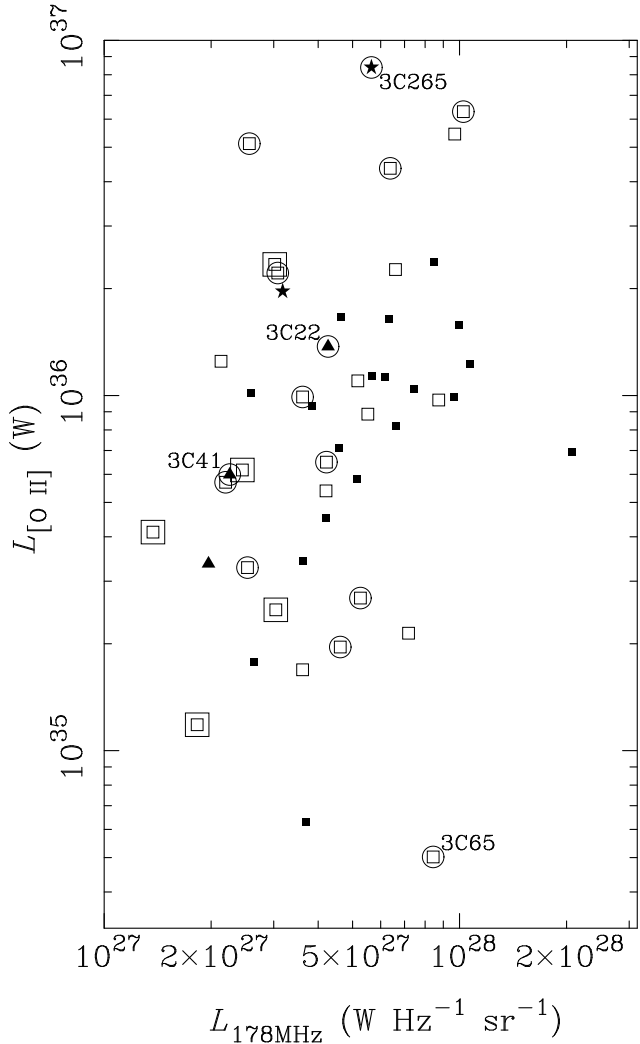}
\caption[]{[O\II]~$\lambda$3727 narrow-line
luminosity versus 178\,MHz radio luminosity for objects from the LRL
sample with $0.65 \leq z < 1.20$ plus the additional UKIRT targets.
Symbols have the same meaning as in Fig.~\ref{fig:cores}.  Emission
line data are taken from the compilation of Jackson \& Rawlings (1997)
with a few updates reported by Willott et al.\ (1998c); six objects
without [O\II] measurements have had their [O\III] luminosities scaled
according to the typical line ratios in McCarthy (1993); an [O\II]
rest-frame equivalent width of 10\,\AA\ has been assumed for the seven
quasars without spectrophotometry, and four objects (3C~220.3, 3C~292,
3C~318 and 3C~325) have not been plotted due to the lack of any
emission line data.}
\label{fig:lines}
\end{figure}

We next consider possible biases introduced by our incomplete UKIRT
observational programme.  The summary numbers included in
Fig.~\ref{fig:cores} show that we have obtained UKIRT data for about
half of the possible target LRL radio galaxies with no significant
redshift bias across the full range.  There are insufficient data to
consider any more than one redshift bin in our later analysis, but we
note that on the basis of Fig.~\ref{fig:cores} there is very little
evidence for any change in quasar fraction with redshift across this
bin (although see Lawrence 1991). Again because of lack of data we are
forced to include the non-LRL objects in our analysis without
considering them separately from the LRL objects. There is some worry
that this might introduce a small bias because these objects were
typically excluded from LRL because their accurate radio flux
densities fall slightly below the LRL limit, and as discussed in
Section~\ref{sec:intro} their lower radio luminosities might be
expected to translate into lower quasar luminosities.  To assess the
impact of this effect we plot in Fig.~\ref{fig:lines} the [O\II]
narrow-line luminosity, $L_{\rm[O~II]}$, versus 178-MHz radio
luminosity, $L_{178}$, for the parent LRL sample plus our additional
UKIRT targets. This plot is relevant because according to unified
schemes the [O\II] luminosity should be much more closely related to
the underlying quasar luminosity than $L_{178}$ (e.g., Hes, Barthel \&
Fosbury 1993). Although Simpson (1998) has suggested that the optical
[O\III] doublet provides a more reliable probe of the intrinsic quasar
luminosity, the wavelengths to which these lines are redshifted for
sources in our sample mean there exist very few measurements of their
strengths. The effects of using different narrow lines are subtle
compared to the points we shall make on the basis of
Fig.~\ref{fig:lines}.  Any correlation in Fig.~\ref{fig:lines} is much
weaker than the one seen when objects spanning several orders of
magnitude in $L_{178}$ are plotted (e.g., Willott et al.\ 1998c). This
demonstrates that the scatter in the $L_{\rm[O~II]}$, $L_{178}$
relation is sufficiently large that the factor $\sim 2$ difference in
$L_{178}$ between the 3CR and LRL objects has negligible effect on the
expected difference in underlying quasar luminosities.
Kolmogorov--Smirnov tests show that the distribution of
$L_{\rm[O~II]}$ for the UKIRT targets is indistinguishable from that
for the unobserved LRL galaxies alone, and from that for the
unobserved LRL galaxies and quasars combined.  The lower scatter
displayed in $L_{\rm[O~II]}$ by the quasars is at least in part due to
that fact that for a significant number of them (7/19) we have
inferred the line luminosity from broad-band photometry by assuming a
single rest-frame equivalent width, whereas in reality there is a
significant spread in this property. We conclude, and shall assume
hereafter, that our UKIRT targets form an unbiased selection of 3C
radio galaxies in the target redshift range.

In Section~\ref{sec:discussion} we will need to consider an unbiased
sample of radio sources including both quasars and radio galaxies. To
do this we must estimate the number of quasars which would exist in a
sample with the same incomplete selection criteria as our radio galaxy
sample. Excluding 3C~454.3, whose presence in LRL is due to
Doppler-boosted core emission leaves 19 LRL quasars with $0.65 \leq z
< 1.20$, compared to 29 galaxies. This quasar fraction (40 per cent)
corresponds to a torus opening angle $\theta_{\rm c} = 53^\circ$.
Our sample of 19 radio galaxies observed with UKIRT (including five
from outside LRL, but which we have argued are not significantly
different) should therefore be accompanied by a `virtual' sample of
12.5 quasars.

\section{Imaging observations}

All objects were observed at UKIRT using the IRCAM3 infrared array
with a nominal scale of 0\farcs281\,pixel$^{-1}$. Our observing
strategy consisted of dithering the telescope to place the object in
nine different locations on the detector array, and after applying a
linearity correction to each frame, we reduced the data in the
following manner. First, each of the nine frames was dark-subtracted,
and all the frames were then scaled to have the same median pixel
value and median-filtered to produce a flat field. This was normalized
and then divided into the individual dark-subtracted frames. These
frames were registered at $J$ and $K$ by determining the individual
image offsets from the centroids of the brightest objects in the
fields where possible, and otherwise (e.g.\ at $L'$, where the high
thermal background and smaller field of view made it impossible to see
anything in the individual images) by using the telescope offsets
recorded in the image headers. Each set of nine images was then
averaged together (ignoring known bad pixels), and objects identified
in this image as contiguous groups of pixels whose values were above a
certain threshold determined from the sky level and noise. A new flat
field was then constructed, ignoring the regions in the individual
frames which contained objects, and the above procedure was
repeated. Usually we took more than one set of nine images, and the
final image was produced by averaging these sets, after additional
registration if necessary. Flux calibration was usually performed by
observing several UKIRT flux standards during the course of each night
and determining the atmospheric extinction coefficient and array
sensitivity in each filter. The $L'$ image of 3C~22 was calibrated
using observations of the single standard star HD~225023 taken
immediately before and after the radio galaxy images.

Our initial observing run of UT 1995 March 23--26 was beset by
problems with the IRCAM3 control software, ALICE. Not only did this
result in a significant loss of observing time, but our $L'$ images
frequently showed residual gradients and other structure due to an
unstable bias level. We excluded from the final coaddition those
images which were most seriously affected by this problem, but in many
cases the images could be reduced in the normal manner and the only
effect was an increased noise level which caused the overall
signal-to-noise ratio to increase more slowly than the square root of
the total integration time. For this reason, our sensitivity at $L'$
during this run was lower than during later runs (where this problem
was not experienced), even though a number of sources received similar
total exposure times. We present our complete observing log in
Table~\ref{tab:log}.

\begin{table}
\caption[]{Observing log. Listed for each filter and each source are
the UT date of observation (YYMMDD) and total useful exposure time (in
seconds). $K$-band imaging of 3C54 was accidentally excluded from the
observing programme.}
\label{tab:log}
\setlength{\tabcolsep}{5pt}
\begin{tabular}{lcc@{\hspace*{6pt}}rc@{\hspace*{6pt}}rc@{\hspace*{6pt}}r}
Name & $z$ & \multicolumn{2}{c}{$J$} & \multicolumn{2}{c}{$K$} &
\multicolumn{2}{c}{$L'$} \\
\hline
3C~22   & 0.935 & 970826 &  540 & 970826 &  540 & 970130 &  540 \\
3C~34   & 0.690 & 970825 & 1080 & 970825 & 1080 & 970825 & 3510 \\
3C~41   & 0.794 & 961104 & 1080 & 961104 &  900 & 961104 & 3240 \\
3C~44   & 0.660 & 970824 & 1080 & 970824 & 1080 & 970825 & 3240 \\
3C~54   & 0.827 & 970826 & 2160 &        &      & 970826 & 3510 \\
3C~55   & 0.735 & 970824 & 1080 & 970824 & 1080 & 970824 & 3510 \\
3C~65   & 1.176 & 961104 & 1080 & 961104 & 1080 & 961104 & 3240 \\
3C~107  & 0.785 & 970824 & 1080 & 970824 & 1080 & 970826 & 3510 \\
3C~114  & 0.815 & 970824 & 1080 & 970824 & 1080 & 970825 & 3240 \\
3C~217  & 0.898 & 950324 &  810 & 950324 &  810 & 950324 & 1820 \\
3C~226  & 0.818 & 950323 &  810 & 950323 &  810 & 950323 & 1960 \\
3C~265  & 0.811 & 950323 &  810 & 950323 &  810 & 950324 & 2630 \\
3C~272  & 0.944 & 950323 &  810 & 950323 &  810 & 950324 & 2540 \\
3C~289  & 0.967 & 950325 &  810 & 950325 &  810 & 950325 & 2120 \\
3C~340  & 0.775 & 970825 & 1080 & 970825 & 1080 & 970825 & 3510 \\
3C~352  & 0.807 & 950325 &  810 & 950325 &  810 & 950325 & 1980 \\
3C~356  & 1.079 & 970825 & 4860 & 950323 & 4860 & 970824 & 3660 \\
3C~368  & 1.132 & 970914 & 1080 & 970914 & 1080 & 970914 & 3510 \\
3C~441  & 0.707 & 970824 & 1080 & 970824 & 1080 & 970824 & 3390 \\
\hline
\end{tabular}
\end{table}

Of those objects observed during the 1995 run, only 3C~265 was
detected, and was extended over several arcseconds east--west. Since
this is not a preferred axis of the host galaxy or radio source, we
suspected that the elongation may have been caused by telescope drift
during the two hours over which our observation was made. This was
confirmed by our 3C~356 images, which were taken over approximately
two hours of real time, and therefore allowed us to measure the true
image offsets from sources on the frames and compare them to the
nominal telescope values. A drift of $\sim 0.9$\,arcsec\,hr$^{-1}$ in
an easterly direction was revealed, and the 3C~265 data were
re-reduced with this drift accounted for. The new image revealed a
bright source, this time only slightly elongated in an east--west
direction. We attribute this elongation to imperfect correction of the
telescope drift and believe the source to be unresolved. We re-reduced
the other $L'$ datasets incorporating the telescope drift correction,
but failed to detect any of the remaining objects. The addition of the
fast guider and tip-tilt secondary mirror for the later observations
meant that those images were not affected in this manner.

\section{Results and analysis}
\label{sec:ir}

\begin{table*}
\caption[]{Aperture photometry (converted to zero airmass) from our
UKIRT images. All limits are $2.5\sigma$. Sources marked $^*$ are not
members of LRL, only 3CR.}
\label{tab:photom}
\begin{tabular}{lcrrr@{ }lrrr@{ }l}
Name & $z$ & \multicolumn{1}{c}{$J$ (3\arcsec)} &
\multicolumn{1}{c}{$K$ (3\arcsec)} & \multicolumn{2}{c}{$L'$
(3\arcsec)} & \multicolumn{1}{c}{$J$ (6\arcsec)} &
\multicolumn{1}{c}{$K$ (6\arcsec)} & \multicolumn{2}{c}{$L'$
(6\arcsec)}
\\ \hline
3C~22  & 0.935 & $17.71\pm0.02$ & $15.74\pm0.03$ & 13.34 & $\pm0.14$
& $17.47\pm0.03$ & $15.57\pm0.03$ & 13.23 & $\pm0.21$ \\
3C~34  & 0.690 & $18.53\pm0.02$ & $16.66\pm0.02$ & $>15.72$ &
& $18.05\pm0.03$ & $16.20\pm0.03$ & $>14.97$ & \\
3C~41  & 0.794 & $18.38\pm0.06$ & $16.17\pm0.05$ & 13.71&$\pm0.13$
& $17.93\pm0.06$ & $15.87\pm0.05$ & 13.49 & $\pm0.20$ \\
3C~44$^*$ & 0.660 & $18.24\pm0.03$ & $16.51\pm0.03$ & $>15.75$ &
& $17.75\pm0.03$ & $16.09\pm0.03$ & $>15.00$ & \\
3C~54$^*$ & 0.827 & $18.69\pm0.03$ & & $>15.60$
& & $18.20\pm0.03$ & & $>14.85$ & \\
3C~55  & 0.735 & $18.92\pm0.04$ & $17.69\pm0.04$ & $>15.84$ &
& $18.57\pm0.05$ & $17.40\pm0.06$ & $>15.09$ & \\
3C~65  & 1.176 & $19.36\pm0.07$ & $17.25\pm0.05$ & 15.53 & $\pm0.49$
& $18.90\pm0.08$ & $16.83\pm0.06$ & $>14.80$ & \\
3C~107$^*$ & 0.785&$19.24\pm0.05$ & $17.36\pm0.04$ & $>15.57$ &
& $18.95\pm0.07$ & $17.13\pm0.05$ & $>14.82$ & \\
3C~114$^*$ & 0.815&$19.24\pm0.05$ & $16.95\pm0.03$ & $>15.72$ &
& $18.75\pm0.06$ & $16.46\pm0.03$ & $>14.97$ & \\
3C~217 & 0.898&$19.12\pm0.04$ & $18.14\pm0.07$ & $>15.32$ &
& $18.75\pm0.05$ & $17.68\pm0.09$ & $>14.57$ & \\
3C~226 & 0.818&$18.88\pm0.05$ & $17.08\pm0.04$ & $>15.41$ &
& $18.39\pm0.06$ & $16.76\pm0.05$ & $>14.66$ & \\
3C~265 & 0.811&$18.09\pm0.04$ & $16.67\pm0.03$ & 14.04 & $\pm0.13$
& $17.66\pm0.04$ & $16.35\pm0.04$ & 13.94 & $\pm0.22$ \\
3C~272$^*$ & 0.944&$19.06\pm0.05$ & $17.14\pm0.04$ & $>15.39$ &
& $18.88\pm0.06$ & $16.86\pm0.05$ & $>14.64$ & \\
3C~289 & 0.967&$18.94\pm0.04$ & $17.31\pm0.05$ & $>15.47$ &
& $18.34\pm0.04$ & $16.79\pm0.06$ & $>14.72$ & \\
3C~340 & 0.775 & $18.94\pm0.03$ & $17.25\pm0.03$ & $>15.70$ &
& $18.64\pm0.05$ & $17.12\pm0.05$ & $>14.95$ & \\
3C~352 & 0.807 & $18.69\pm0.03$ & $17.32\pm0.05$ & $>15.41$ &
& $18.06\pm0.03$ & $16.84\pm0.06$ & $>14.66$ & \\
3C~356a& 1.079 & $19.34\pm0.03$ & $17.57\pm0.03$ & $>15.65$ &
& $18.92\pm0.03$ & $17.18\pm0.04$ & $>14.90$ & \\
3C~356b& 1.079 & $19.40\pm0.03$ & $17.47\pm0.03$ & $>15.65$ &
& $18.74\pm0.03$ & $16.86\pm0.03$ & $>14.90$ & \\
3C~368$^\dagger$& 1.132 & $19.91\pm0.18$ & $18.12\pm0.15$ & $>15.79$ &
& $19.26\pm0.29$ & $17.55\pm0.23$ & $>15.04$ & \\
3C~441 & 0.707 & $18.48\pm0.03$ & $16.76\pm0.03$ & $>15.80$ &
& $18.00\pm0.04$ & $16.32\pm0.03$ & $>15.05$ & \\
\hline
\end{tabular}

$^\dagger$We have subtracted the foreground Galactic star and assumed
that the nucleus is at the location identified by Stockton, Ridgway \&
Kellogg (1996).
\end{table*}

In Table~\ref{tab:photom} we present the results of photometry in
3-arcsec and 6-arcsec apertures for the galaxies observed. Although
some of the limits at $L'$ are fairly bright ($L' < 15.5$) due to the
reduced exposure times caused by the problems with the instrument
electronics, most observations were not affected and the limits
provide useful constraints as detailed in Section~\ref{sec:discussion}.

\begin{figure*}
\vspace*{194mm}
\includegraphics{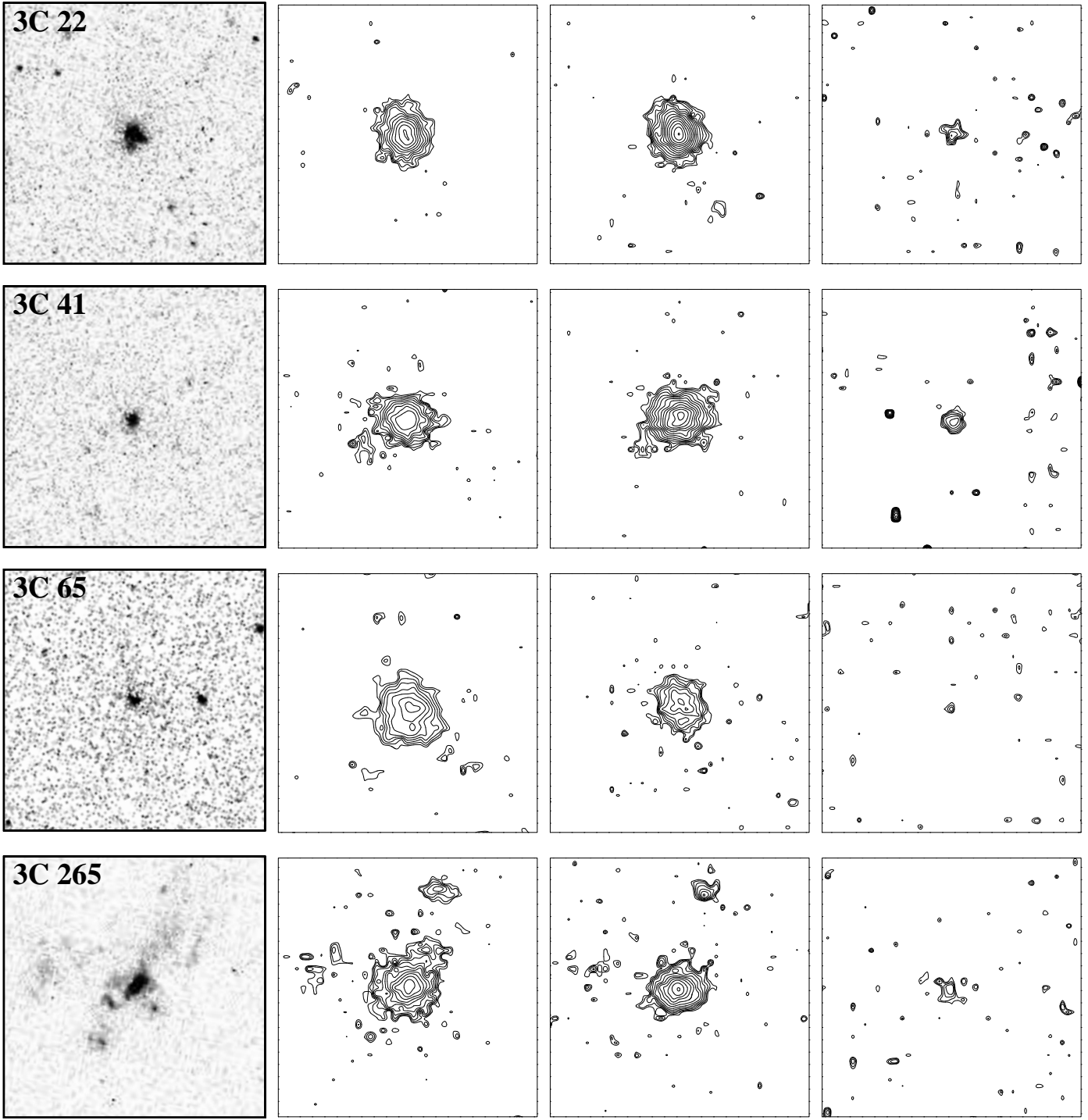}
\caption[]{Images of the four radio galaxies detected at $L'$. The
greyscale is an optical \HST\ image, and the contour maps are, from
left to right, $J$, $K$, and $L'$, with contours spaced at intervals
of 0.25\,mag\,arcsec$^{-2}$. Each field is 12\,arcsec on a side. The
galaxies and the filters in which the \HST\ images were taken are, from
top to bottom, 3C~22 (F622W), 3C~41 (F555W), 3C~65 (F675W), and 3C~265
(F555W).}
\label{fig:images}
\end{figure*}

Aperture photometry of our images at $J$ and $K$ agrees well with that
of Best et al.\ (1998), with an r.m.s.\ difference of 0.10\,mag and no
tendency for our magnitudes to be systematically fainter or
brighter. The agreement with the aperture photometry of Lilly \&
Longair (1984) is also good, although the uncertainties in their
measurements are often large. We note that errors at the level of
0.1\,mag are insufficient to affect our results. Four radio galaxies
were detected at $L'$, and we present our {\it JKL$'$\/} images of
these, together with an optical \HST\ image of each, in
Fig.~\ref{fig:images}.

\begin{figure*}
\includegraphics{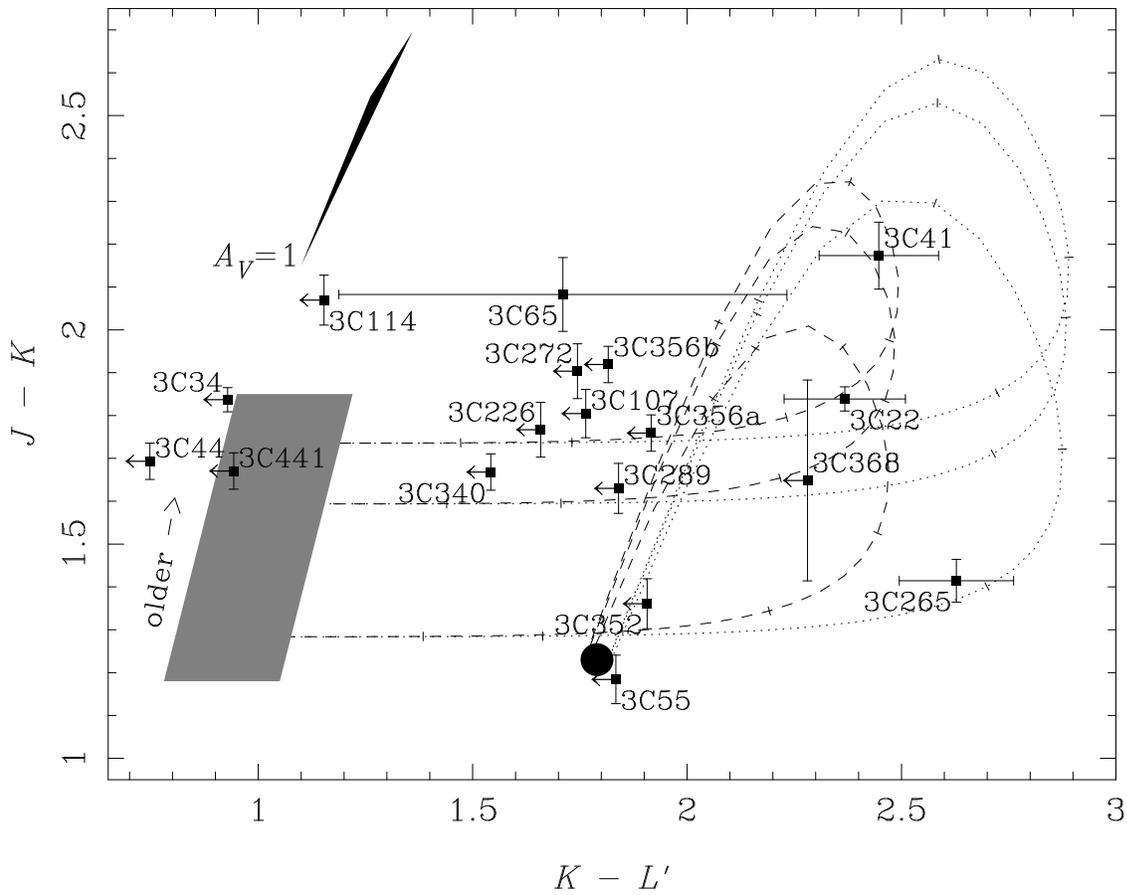}
\vspace*{120mm}
\caption[]{Colour--colour diagram for the galaxies observed, as
measured in 3-arcsec apertures, and corrected for Galactic extinction
using the prescription of Rowan-Robinson et al.\ (1991). The shaded
region on the left indicates the colours of stellar populations formed
in a 1-Gyr burst with ages 1--5\,Gyr, with older populations being
redder in $J-K$ (as indicated). The solid circle represents the
combined colours of an unreddened quasar and stellar population. The
dashed lines indicate the loci of colours for a reddened quasar and an
unreddened stellar population of ages 1,2,5\,Gyr at $z=1$ (there is
little variation in the loci with redshift within the range we are
considering) where the quasar is intrinsically five times brighter
than the host galaxy in the $K$-band. The dotted lines are the loci
for the case where the quasar is ten times brighter than its host
galaxy. Each curve has tick marks at $A_V = 2$,5,10,15,30\,mag. The
wedge in the top left of the diagram shows the effect of one magnitude
of visual extinction at $0.65 \leq z < 1.20$.}
\label{fig:colcol}
\end{figure*}

In Fig.~\ref{fig:colcol} we present our photometry graphically in a
colour--colour diagram. We also plot the loci for combinations of
reddened quasar plus unreddened starlight. This figure can be used to
make an approximate determination of the nature of individual sources;
for example, 3C~22 and 3C~41 are clearly lightly reddened ($A_V
\approx 2$--5) quasars, whereas 3C~34 and 3C~44 appear to be normal
galaxies with no sign of AGN activity in their infrared photometry.
However, some objects, such as 3C~55, are more ambiguous, but this
ambiguity can be removed by considering both the $K$-band morphology
and the $L'$-band magnitude of such sources. Considering 3C~55, it is
more than two magnitudes fainter at $K$ than the locus of 3C quasars
(Fig.~\ref{fig:kz}), and it is clearly resolved in our near-infrared
images, indicating that its putative central quasar does not
contribute significantly at these wavelengths. The same line of
reasoning can be applied to the other radio galaxies whose $K-L'$
upper limits do not apparently exclude them from being
lightly-reddened quasars in Fig.~\ref{fig:colcol}. The case of 3C~65
is rather more ambiguous --- for the present, we shall assume that it
too is lightly-reddened (see also Lacy et al.\ 1995).

Although Fig.~\ref{fig:colcol} can be used to estimate directly the
extinctions to the quasars in those radio galaxies detected at $L'$,
we opt for a less subjective approach. In this method, we adopt
various galaxy SEDs (1-Gyr bursts with ages ranging from 1\,Gyr to the
age of the Universe at the source redshift) and solve for the quasar
and galaxy magnitudes and the nuclear extinction, using our {\it
JKL$'$\/} photometry. As Fig.~\ref{fig:seds} shows, the $J-K$ and
$K-L'$ colours are not especially sensitive to age at $z \sim 1$, so
there is little variation in the derived quantities with the assumed
galaxy age. With the added constraint that the $K$ magnitude of the
host galaxy should lie on the $K$--$z$ relation, we then determine the
range of each parameter. In Table~\ref{tab:intrinsic} we present the
derived nuclear extinctions and unobscured quasar $K$ magnitudes. We
also provide the estimated fraction of $K$-band non-stellar light in a
3-arcsec aperture.

\begin{table}
\centering
\caption[]{Estimated extinctions and unobscured quasar magnitudes, and
estimated fraction of non-stellar emission in a 3-arcsec aperture at
$K$, determined using the method outlined in Section~\ref{sec:ir}.}
\label{tab:intrinsic}
\begin{tabular}{lrcc}
Object & \multicolumn{1}{c}{$A_V$} & $K_0$ & $f_K$ \\
\hline
3C~22  &  $3\pm1\;\;$ & $14.77\pm0.18$ & $0.90^{+0.10}_{-0.12}$ \\
3C~41  &  $6\pm2\;\;$ & $15.06\pm0.21$ & $0.51^{+0.13}_{-0.11}$ \\
3C~65  &  $3\pm1\;\;$ & $17.70\pm0.22$ & $0.28^{+0.12}_{-0.09}$ \\
3C~265 & $15\pm4\;\;$ & $14.31\pm0.45$ & $0.14^{+0.05}_{-0.06}$ \\
\hline
\end{tabular}
\end{table}

\section{Comparison with other data}

\subsection{Re-analysis of \HST\ data}
\label{sec:hst}

Although we have obtained extinction estimates from our infrared data
alone, they are rather uncertain and we would also like to be able to
confirm the correctness of our values. The easiest way to do this is
by extending our study to optical wavelengths, and searching for point
sources there. However, with the exception of 3C~22, which McCarthy
(1988) reported to be unresolved in an optical continuum image, the
quasar nuclei in the other objects are predicted to be too faint in
the optical ($V\simgt24$) to be readily detectable from ground-based
images. We therefore turn to the archived {\it Hubble Space
Telescope\/} WFPC2 observations of these objects (PID 1070, P.I.\
Longair; see Best, Longair \& R\"{o}ttgering 1997) where the
point-source sensitivity is much greater. The calibrated science data
files were retrieved from the \HST\ data archive and each pair of
images combined and cleaned of cosmic rays using the {\sc iraf} task
{\it crrej\/}, with pixels unaffected by cosmic rays in both images
being assigned an exposure-time-weighted average. Flux calibration was
performed using the on-line WFPC2 exposure time calculator, assuming
an $\alpha = 0$ source spectrum.

We re-analysed these data using our own two-dimensional fitting
procedure. Heavy smoothing of the data revealed that a
circularly-symmetric de Vaucouleurs (1948) profile provided an
adequate fit to the host galaxy component for 3C~22, 3C~41 and 3C~65
while no reasonable galaxy model could fit the very peculiar structure
of 3C~265 (see Fig.~\ref{fig:images}) so no fit was attempted. For
the other objects, we fit a model WFPC2 point spread function
(generated with the {\sc tiny tim} software; Krist 1995) and a de
Vaucouleurs host galaxy. The point source flux and the total flux and
effective radius of the galaxy component were varied to minimize the
value of $\chi^2$. The uncertainty assigned to each pixel was
calculated as the Poisson error on the source counts added in
quadrature to the standard deviation of pixel values measured in a
blank region of the chip. Values of $\chi^2$ (and residual images)
were produced as a function of fitting radius to check for systematic
effects (such as incorrect background subtraction), and as none were
found a fitting radius of 15 pixels (1.5 arcsec) was used in all
cases, and fluxes measured over the same aperture.  The residual
images were structureless with the exception of 3C~22 for which the
two companion objects discussed by Best et al.\ (1997) were clearly
revealed (and possibly the F555W image of 3C~41 where there was some
evidence for excess partially-resolved emission near the nucleus); in
the case of 3C~22 the fitting procedure was repeated after masking out
data at the positions of the companions. The quantitative results of
this analysis are presented in Table~\ref{tab:hst}; also listed are
the fluxes estimated using the intrinsic nuclear fluxes and
extinctions listed in Table~\ref{tab:intrinsic}, accounting for the
Galactic extinctions given in Table~\ref{tab:3cseds}. The exclusion of
3C~265 from this analysis is not a great problem because its predicted
transmitted nuclear flux is several orders of magnitude fainter than
the detection limit in the \HST\ images, and spectropolarimetry (Dey
\& Spinrad 1996) suggests that any compact component is more likely to
be light scattered close to the nucleus. The agreement between
predicted and observed \HST\ fluxes is generally very good, with the
exception of the F555W flux of 3C~41, which is predicted to be much
fainter than is observed. We discuss this difference below.

\begin{table*}
\centering
\caption[]{The results of our host galaxy plus nuclear point source
decomposition performed on the archived \HST\ images of the four
objects detected at $L'$, as described in Section~\ref{sec:hst}. Data
obtained on 3C~22 using the F1042M filter have extremely poor
sensitivity and are ignored. For the best-fitting model in each
object/filter combination, we list the reduced $\chi^2$, the effective
radius $r_{\rm e}$ of the host galaxy, the percentage contribution to
the \HST\ counts within a 3-arcsec diameter from the galaxy and
quasar, and the flux density of the quasar component. Note that these
percentages typically sum to less than 100 per cent because other
components are revealed by the modelling process, and evident in the
residual map. Significant differences between our best-fit parameters
and those derived from the same \HST\ dataset by Best et al.\ (1998)
are discussed in Section~\ref{sec:hst}. Also listed is the flux
density of the quasar nucleus estimated from our near-infrared data
(see Section~\ref{sec:ir}).}
\label{tab:hst}
\begin{tabular}{llccrrrl}
Name & WFPC2 & $\chi^2_\nu$ & $r_{\rm e}$ & \% gal & \% QSO &
\multicolumn{2}{c}{Fluxes ($\mu$Jy)} \\
 & Filter & & (arcsec) & & & Observed & Estimated \\ 
\hline
3C~22 & F622W  & 3.3 & 0.8 &   80  &   15  &  $1.8$ & $3.01^{+8.04}_{-2.19}$\\
3C~41 & F555W  & 2.0 & 0.6 &   60  &   30  &  $1.3$ & $0.03^{+0.13}_{-0.02}$\\
      & F785LP & 2.5 & 0.8 &   90  &   10  &  $2.8$ & $0.89^{+2.12}_{-0.52}$\\
3C~65 & F675W  & 1.3 & 1.2 & $>90$ & $<10$ & $<0.2$ & $0.33^{+0.76}_{-0.23}$\\
      & F814W  & 1.0 & 1.5 & $>90$ & $<10$ & $<0.7$ & $0.74^{+1.27}_{-0.47}$\\
\hline
\end{tabular}
\end{table*}

It is worth noting some differences in our fitting results to those
obtained previously by Best et al.\ (1998) using a one-dimensional
fitting algorithm on the same dataset. While we obtain similar results
for 3C~65, our effective radii for 3C~22 and 3C~41 are much smaller
than those of Best et al. We suggest that these authors have
overestimated the radii in these galaxies because their bright nuclei
force them to exclude all data within a radius of 1 arcsec, and the
fit is therefore dominated by the very tenuous detection of the
low-surface brightness extended starlight --- their outermost
datapoints correspond to fluxes per pixel well below one-tenth of the
sky noise. At these levels, one cannot remove all the companion
objects, and there will also be low-energy cosmic rays left in the
final image, with the result that the surface brightness will be
overestimated, leading in turn to an overestimate of the effective
radius. Fitting the host and nucleus simultaneously using the full
two-dimensional data alleviates all of these concerns, allowing data
from the central regions to constrain $r_{\rm e}$ as well as giving a
very low weight to individual discrepant pixels at large radii.

\begin{figure*}
\vspace*{50mm}
\includegraphics{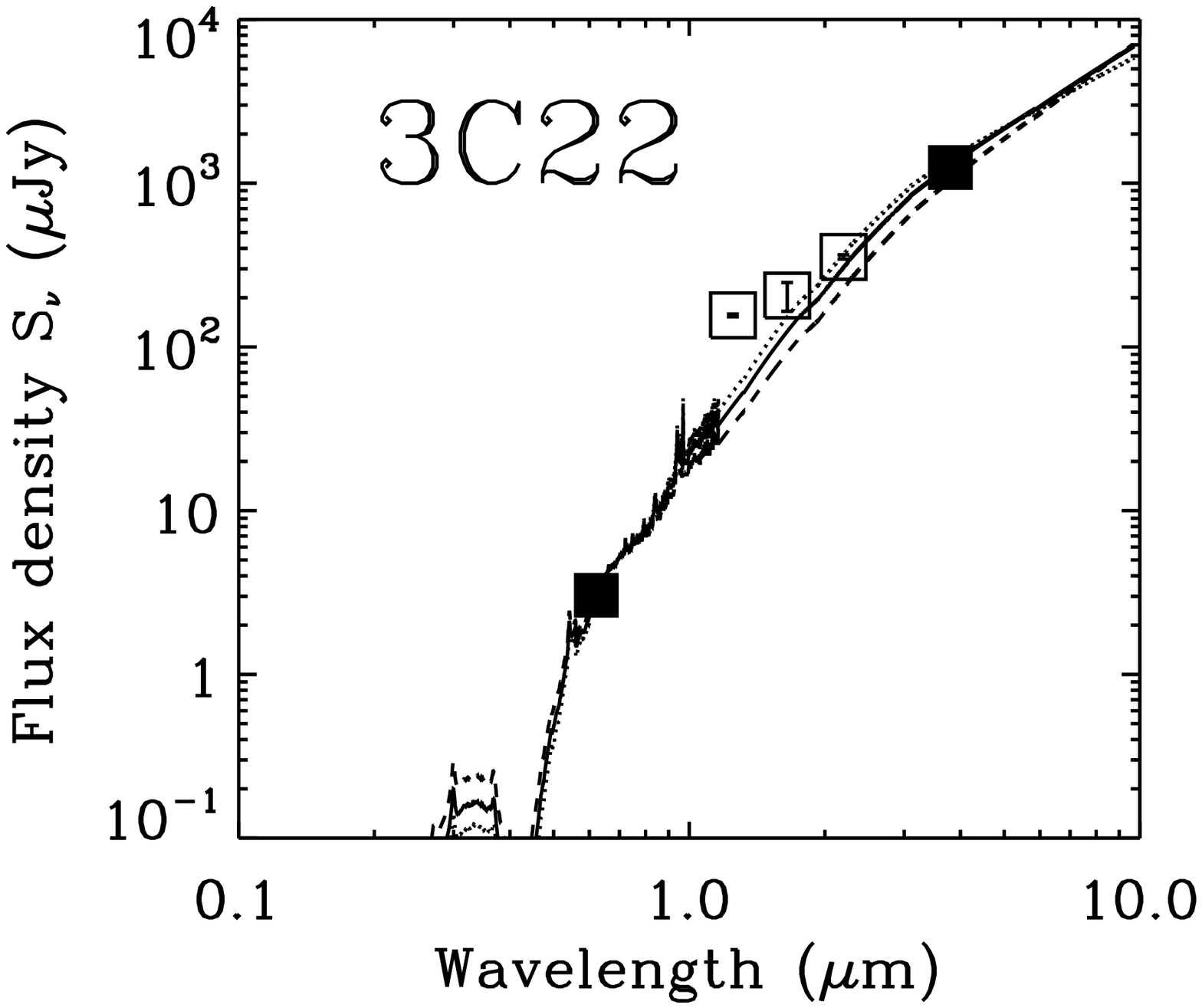}
\includegraphics{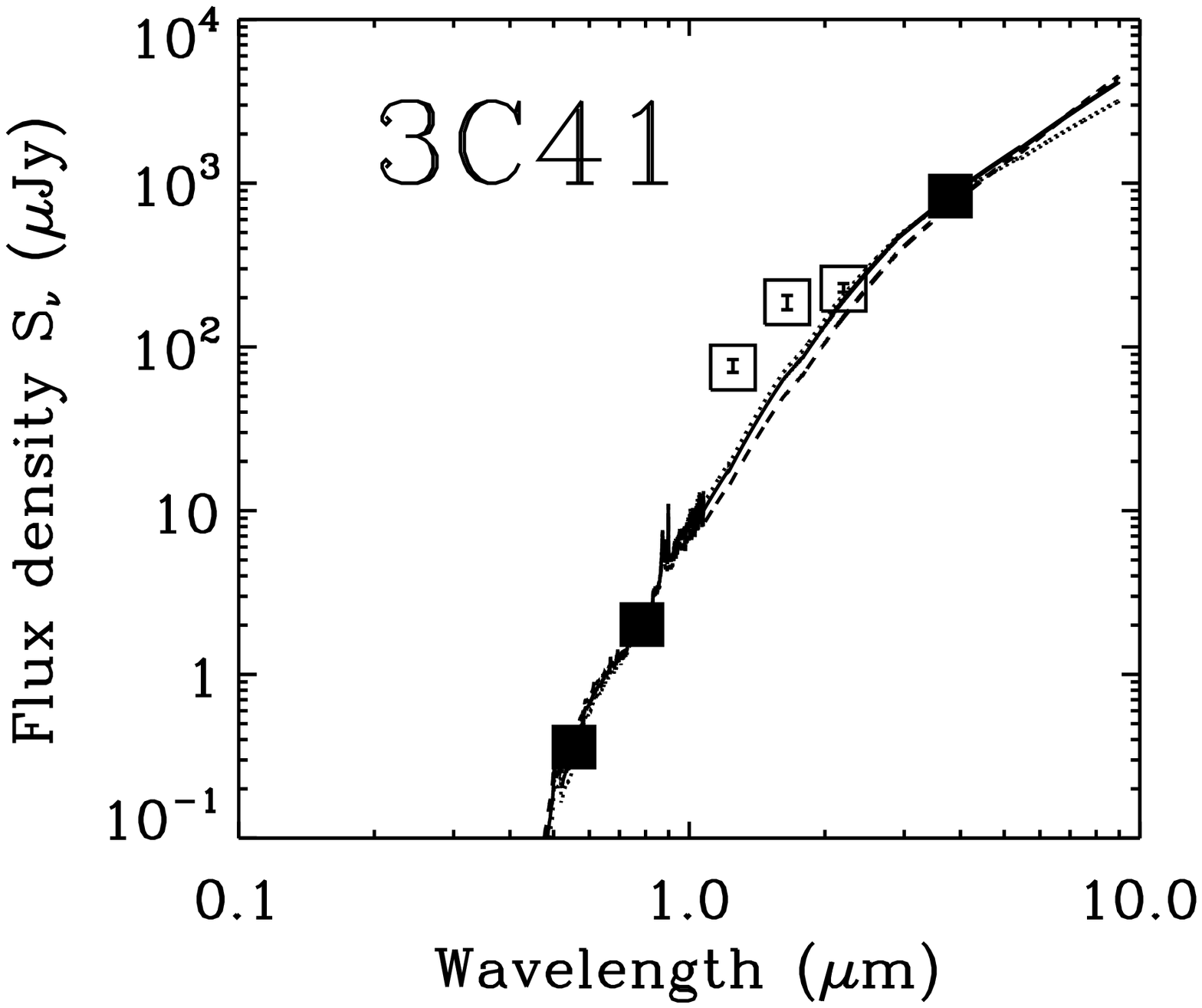}
\includegraphics{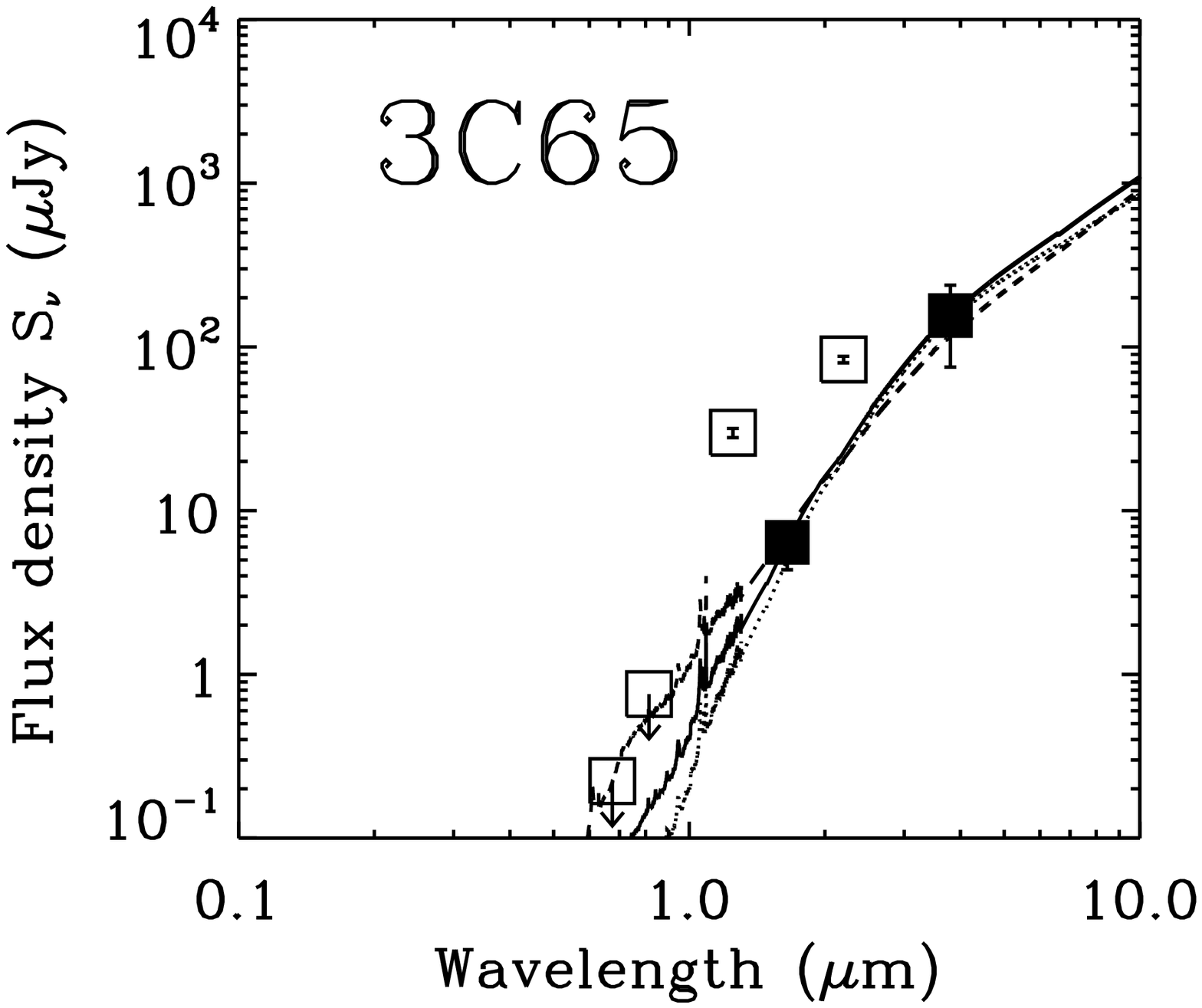}
\caption[]{Spectral energy distributions (SEDs) for 3C~22, 3C~41, and
3C~65. The data have been corrected for extinction in the Milky Way
using the extinction values given in Table~\ref{tab:3cseds}, and are
taken from this paper, Lilly \& Longair (1984) and, in the case of
3C~65, Rigler \& Lilly (1994). Filled data symbols represent points
where we have assumed that all the flux is transmitted light from the
quasar nucleus.  The full line represents a fit using the canonical
quasar spectra reddened using a MW-type extinction curve (full
lines). The dotted line is the most-reddened quasar fit consistent
with the data allowing the intrinsic SED for the quasar to be bluer by
subtracting 0.3 from the spectral index of both the power-law
components discussed in Section~\ref{sec:intro}, and the dashed line
the least-reddened quasar fit consistent with the data with a redder
intrinsic SED (produced by adding 0.3 to both spectral indices).
These assumptions about the spread in the intrinsic SEDs are based on
the quasar SED study of Neugebauer et al.\ (1987).}
\label{fig:3cseds}
\end{figure*}

After using the \HST\ data to isolate compact components in the
optical it was possible to make more detailed studies of the SEDs, and
hence better estimates of the nuclear extinctions of our UKIRT $L'$
detections. A minimum-$\chi^{2}$ fitting procedure was used to find
the most likely values of $A_{V}$, and the fractions of nuclear flux
in 3-arcsec apertures at $J$, $H$ and $K$. In performing this
analysis, we assumed that the $L'$ and \HST\ point source fluxes are
dominated by transmitted light from the reddened quasar nucleus,
except for 3C~41 where we make a correction for scattered light
described below. The results of this analysis are presented in
Fig.~\ref{fig:3cseds} and Table~\ref{tab:3cseds}. We estimated errors
on the fitted quantities by varying the intrinsic colour of the quasar
over the range observed in naked quasars accounting properly for the
photometric error on each (assumed) nuclear flux data point (see
Fig.~\ref{fig:3cseds}); there were no significant differences in the
derived quantities from fits which used LMC- and SMC-type dust
extinction curves (from Pei 1992). Note the extremely good agreement
between the extinction estimates, and especially the fractions of
non-stellar $K$-band light, determined from the infrared data alone
(Table~\ref{tab:intrinsic}) and the more refined values incorporating
the \HST\ data (Table~\ref{tab:3cseds}). This indicates the ability of
infrared imaging alone to provide quantitative, as well as
qualitative, statements about the nature of radio galaxies. Throughout
the remainder of this paper, we adopt the extinction estimates listed
in Table~\ref{tab:3cseds} in preference to those of
Table~\ref{tab:intrinsic}. Brief notes on an object-by-object basis
follow.

\begin{table*}
\caption[]{Results of our fits to the observed spectral energy
distributions of 3C~22, 3C~41, and 3C~65: the nuclear extinction and
fractional contributions of the quasar nucleus in the {\it JHK\/}
infrared bands in 3-arcsec apertures are listed. The Galactic
extinction, $A_V$ (MW), has been estimated from the {\it IRAS\/}
100\,$\mu$m maps obtained through {\it SkyView\/} (McGlynn, Scollick
\& White 1996) and the prescription of Rowan-Robinson et al.\
(1991). Note a typographical error in the value of the Galactic
extinction for 3C~22 listed in the caption to Figure~3 of Rawlings et
al.\ (1995).}
\label{tab:3cseds}
\begin{tabular}{lccclc}
Name & $A_V$ (MW) & $A_V$ & $f_J$ & \multicolumn{1}{c}{$f_H$} & $f_K$ \\
\hline
3C~22 & 0.66 & $3.6\pm0.3$ & $0.30\pm0.10$ & $0.70\pm0.20$ & $0.85\pm0.15$ \\
3C~41 & 0.20 & $4.4\pm0.4$ & $0.25\pm0.05$ & $0.35\pm0.10$ & $0.75\pm0.20$ \\
3C~65 & 0.15 & $4.8\pm1.3$ & $0.06\pm0.05$ & $0.10\ddag$   & $0.25\pm0.05$ \\
\hline
\end{tabular}

\ddag Following Rigler \& Lilly (1994) we have assumed that 
10\% of the $H-$band flux measured by Lilly \& Longair (1984) lies in
a nuclear component arising from reddened transmitted quasar light.
\end{table*}

\subsection{Notes on individual objects}

\begin{description}

\item[{\bf 3C~22}]
A seemingly straightforward case of a nucleus reddened by $A_{V} = 3.6
\pm 0.3$ which contributes $\sim 80$ per cent of the 3-arcsec
aperture $K$-band light, and dominates the unresolved component in the
\HST\ data. The lack of optical polarization (see Leyshon \& Eales
1998) supports this picture. The extended near-infrared light is
presumably from the host galaxy which is clearly visible in the \HST\
image.

\item[{\bf 3C~41}]
As we mentioned in Section~\ref{sec:hst}, the observed optical fluxes
of the compact nucleus are much larger than we inferred from our
infrared imaging, suggesting the presence of an additional
component, such as scattered light from the nucleus. This is supported
by the observed polarization properties of 3C~41 reported in Leyshon
\& Eales (1998), where the $V$-band polarization is $P = 9.3 \pm
2.3$\%. For electron-scattering over a small range of angles about a
mean angle, $\theta$, the intrinsic polarization of electron-scattered
radiation is $P = 100\% \times (1 - \cos^2 \theta) / (1 + \cos^2
\theta)$, and as a lightly-reddened quasar, the viewing angle should
be only slightly larger than the critical angle of the torus,
$\theta_{\rm c} = 53^\circ$. Therefore the intrinsic polarization of
the scattered light should be around 50\%, and approximately one-fifth
of the total F555W flux (i.e., $\sim 1\,\mu$Jy) would need to be
scattered to produce the observed fractional polarization. This
fraction may be larger if the observed polarization has been diluted
by virtue of being averaged over a range of scattering angles, or if
the scattering medium is dust. If the scatterers have an asymmetric
distribution within the illumination cones, our estimate of the
scattering angle will be in error, although this could serve to either
increase or decrease the intrinsic polarization. In order to perform a
quantitative analysis, however, we assume that this is the true
contribution of scattered light. The contribution from scattered light
in the F785LP filter is predicted to be similar (from Fig.~4 of
Leyshon \& Eales 1998), and we therefore fit the SED after subtracting
1\,$\mu$Jy from each of the two \HST\ point source fluxes to account
for the scattered component. We find $A_V = 4.4 \pm 0.4$ and a
transmitted quasar light contribution of $\sim 85$\% to the 3-arcsec
aperture $K$-band emission. As was the case for 3C~22, the extended
near-infrared light can be ascribed to the host galaxy mapped in the
\HST\ image.

\item[{\bf 3C~65}] 
There are two reasons why it remains uncertain whether we have
detected non-stellar emission at $L'$ in this object.  First, the
detection itself is significant only at the $2.7\sigma$ level and then
only in the 3-arcsec aperture. Secondly, the separation between the
location of 3C~65 and possible loci of stellar populations in
Fig.~\ref{fig:colcol} is only marginally significant. However, both
Lacy et al.\ (1995) and Stockton, Kellogg \& Ridgway (1995) have
argued previously for the presence of a quasar nucleus, and most
importantly for our purposes an analysis by Rigler \& Lilly (1994) of
a 0.6-arcsec resolution $H$-band image points to a nuclear component
contributing $\sim 10$\% of the flux.  Our SED fit uses this point,
the $L'$ detection and the \HST\ limits to suggest that there is
indeed a quasar present which has $A_{V} = 4.8 \pm 1.3$. As suggested
previously by Lacy et al.\ (1995), 3C~65 seems seems have an unusually
faint quasar nucleus for a powerful radio source. In the absence of
reddening we estimate that it would have $M_{B} = -24.8$, 1.5\,mag
fainter than the intrinsic luminosity of the 3C~22 quasar and 0.6\,mag
fainter than the 3C~41 quasar. Note that 3C~65 is a prominent outlier
in Fig.~\ref{fig:lines} in the sense that its narrow emission line
luminosity lies factors of 26 and 11 below the line luminosities of
3C~22 and 3C~41, respectively.

\item[{\bf 3C~265}]
Although the complex optical structure of 3C~265 excludes an analysis
of the {\it HST\/} data, a visual inspection of the images indicates
that no point source is present at the location of the nucleus,
supporting our higher extinction estimate of $A_V \approx 15$ for this
source. The fact that 3C~265 is convincingly detected at $L'$ despite
such large extinction is undoubtedly due to the very luminous quasar
it contains, as evidenced by its line luminosity
(Fig.~\ref{fig:lines}). Although this source displays strong, extended
optical polarization (Jannuzi \& Elston 1991), the near-infrared
polarization is lower ($\sim 5$\% in the $H$-band; Jannuzi, private
communication). The thermal-infrared polarizarion is likely to be even
lower, since the overall SED of 3C~265 rises more rapidly in the
thermal-IR than a quasar SED. Even assuming electron scattering, the
scattered nuclear component cannot contribute more than $\sim (3/f)$\%
of the $L'$-band flux, where $f$ is the fractional polarization of the
scattered component, unless it too is heavily reddened. We are
therefore confident that our $L'$ detection of this source represents
the true, transmitted quasar continuum.
\end{description}

\subsection{H$\alpha$ spectroscopy}

\begin{figure}
\includegraphics{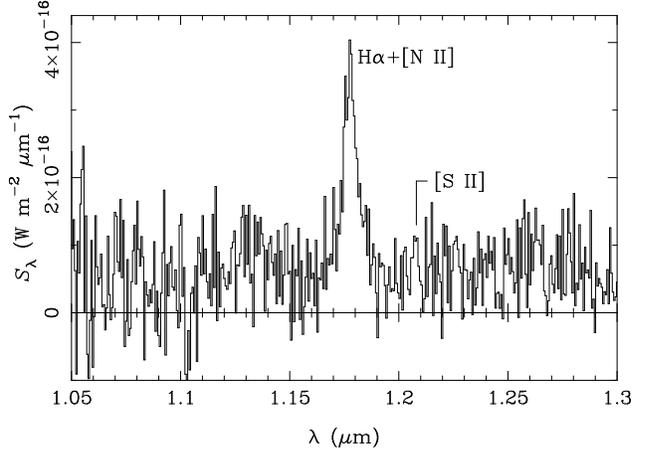}
\vspace*{60mm}
\caption[]{$J$-band spectrum of 3C~41. Note the broad base to the
H$\alpha$+[N\II] complex and the apparent detection of [S\II].}
\label{fig:j3c41}
\end{figure}

As we mentioned in Section~\ref{sec:intro}, a low value of the nuclear
reddening may be indicated by broad wings on the H$\alpha$ emission
line, which is redshifted into the $J$ window for 3C~22, 3C~41 and
3C~265 (but lies where the atmosphere is opaque for 3C~65). Economou
et al.\ (1995) and Rawlings et al.\ (1995) have both shown the
existence of broad H$\alpha$ in 3C~22, and inferred low values of the
extinction ($A_V < 4$ and $A_V = 2 \pm 1$, respectively), in agreement
with our own estimate. Given the relatively low line of sight
extinction derived for the quasar within 3C~41, we took a $J$-band
spectrum with CGS4 on UKIRT on the night of UT 1997 August 12 to look
for broad H$\alpha$. We used the standard `ABBA' technique (e.g.,
Eales \& Rawlings 1993) for observing the source, keeping it on the
array at all times so that all 48\,minutes of integration contributed
to the object signal. The 75\,lines\,mm$^{-1}$ grating was used in
second order with the short camera and a 1-pixel (1.22-arcsec) slit to
provide a resolving power of $\lambda/\Delta\lambda \approx 1000$ at
the wavelength of H$\alpha$. We used $2\times2$ oversampling and the
spectra were interleaved with bad pixel masking using the {\sc CGS4DR}
software package before being exported to {\sc iraf} for coaddition,
sky-subtraction, and extraction of the one-dimensional spectra. The
final reduced spectrum (extracted from a 1.22-arcsec square aperture)
is presented in Fig.~\ref{fig:j3c41}. Broad wings are clearly visible
on the H$\alpha$ line. The convolution of this spectrum with the UKIRT
$J$ filter gives a synthetic magnitude $J=19.13$, which agrees well
with the value $J=19.29$ determined from our image, given that the
image was taken in poorer seeing.

\begin{figure}
\includegraphics{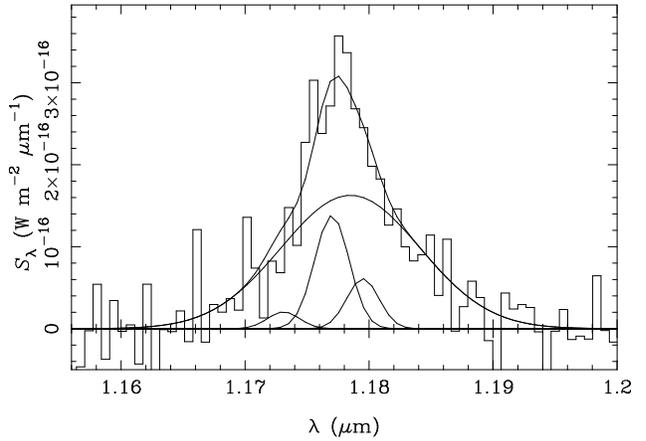}
\vspace*{60mm}
\caption[]{Continuum-subtracted $J$-band spectrum of 3C~41 and best fit of
Gaussian lines, as described in the text.}
\label{fig:j3c41fit}
\end{figure}

We have analysed the spectrum of 3C~41 to derive an independent
extinction estimate. First a fit was made to the continuum using ten
cubic spline pieces and subtracted from the data. We then used the
{\sc specfit} package (Kriss 1994) to fit the H$\alpha$+[N\II] blend
with a combination of four Gaussians, representing the three narrow
lines and a broad H$\alpha$ component. We fixed the relative
wavelengths and intensities of the [N\II] lines and demanded that they
have the same width, but allowed the other parameters to vary
freely. The results are presented in Fig.~\ref{fig:j3c41fit} and
Table~\ref{tab:j3c41fit}.

The results of our line fitting are very similar to those found by
Economou et al.\ (1995) for 3C~22, namely: a rest-frame equivalent
width of 470\,\AA\ for the blend ({\it cf.\/} 540\,\AA\ for 3C~22); a
relative redshift of 1050\,km\,s$^{-1}$ (960\,km\,s$^{-1}$) for the
broad H$\alpha$ line with respect to the narrow line; and ratios
[N\II]~$\lambda$6584/H$\alpha = 0.56\pm0.17$ ($0.59\pm0.14$) and
H$\alpha_{\rm b}$/H$\alpha_{\rm n} = 4.19\pm0.75$ ($3.92\pm0.49$).
Assuming an intrinsic ratio of broad to narrow H$\alpha$ of 40 (e.g.,
Jackson \& Eracleous 1995), this implies $A_V \approx 3$ as the
differential extinction between the broad and narrow lines, in fair
agreement with our more refined estimate from the spectral energy
distribution.

\begin{table}
\caption[]{Results of four-Gaussian fit to the continuum-subtracted
spectrum of 3C~41. The line widths have not been corrected for the
instrumental resolution of 300\,km\,s$^{-1}$.}
\label{tab:j3c41fit}
\begin{tabular}{lcrr}
Line & Redshift & \multicolumn{1}{c}{Flux} & \multicolumn{1}{c}{FWHM}
\\ & & \multicolumn{1}{c}{($10^{-19}$\,W\,m$^{-2}$)} &
\multicolumn{1}{c}{(km\,s$^{-1}$)} \\
\hline
H$\alpha_{\rm n}$ & $0.7929 \pm 0.0005$ &  $5.06\pm0.81$ &  $771\pm111$ \\
{}[N\II]$^a$ &      $0.7911 \pm 0.0008$ &  $2.83\pm0.74$ &  $695\pm178$ \\
H$\alpha_{\rm b}$ & $0.7964 \pm 0.0010$ & $21.25\pm1.67$ & $3438\pm335$ \\
\hline
\end{tabular}

$^a\lambda$6584 only.
\end{table}

While Economou et al.\ (1998) have also used CGS4 to detect broad
wings on the H$\alpha$ line in 3C~41, their spectrum of 3C~265 does
not reveal a broad line. This non-detection is in agreement with the
much larger extinction we derive to the quasar in this object.

\section{Discussion}
\label{sec:discussion}

\begin{figure}
\vspace*{92mm}
\includegraphics{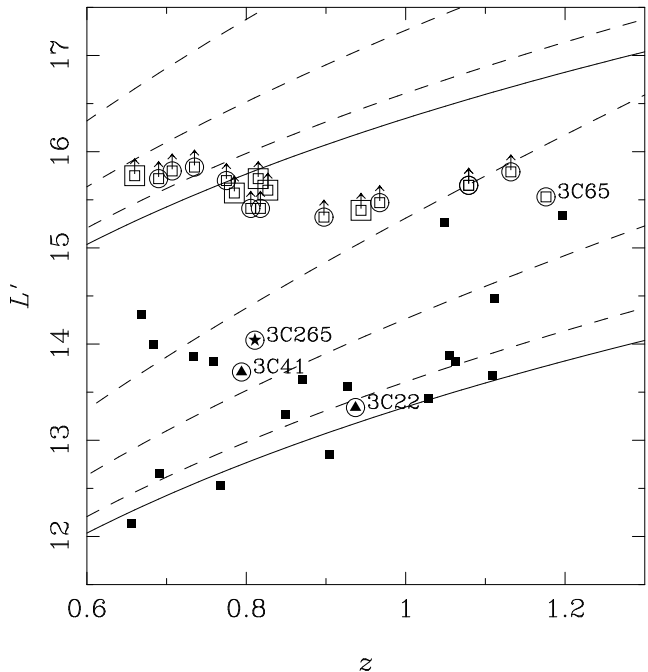}
\caption[]{The $L'$ versus redshift plane for the LRL quasars plus our
sample of radio galaxies; symbols have the same meaning as in
Fig.~\ref{fig:cores}. $L'$ magnitudes for the LRL quasars have been
estimated from the optical magnitudes listed in LRL (except for 3C~190
where LRL list a photographic magnitude, and we elect to use the {\it
HST\/} magnitude of de Vries et al.\ 1997), using the canonical quasar
spectrum to extrapolate to longer wavelengths. We exclude 3C~343 from
this plot since, although classified as a quasar by LRL, it is clearly
resolved by de Vries et al.\ (1997). The solid lines show the loci of
quasars with the canonical spectrum and no reddening for two values of
absolute magnitude ($M_B=-26$ for the lower curve, $M_B = -23$ for the
upper). The dashed lines show the loci of the same quasars reddened
internally by $A_V=2,7$ and $15$.}
\label{fig:lz}
\end{figure}

In Fig.~\ref{fig:lz} we plot the $L'$ magnitudes and limits of our
UKIRT targets against redshift, and compare these data with reddened
quasar spectra. We also plot the estimated $L'$ magnitudes of the LRL
quasars in the same redshift range, extrapolated from optical (usually
$V$-band) magnitudes, assuming zero reddening. Because the
differential reddening between (observed-frame) $V$ and $L'$ is large
(approximately 2\,mag for $A_V=1$), even modest amounts of extinction
will move the quasars up the diagram significantly. Indeed, the two
apparently faint quasars at high redshift are 3C~212 and 3C~190, which
are both intrinsically red (Smith \& Spinrad 1980) and will therefore
be much brighter at $L'$ than we have estimated.

Bearing this in mind, it is clear that our $L'$ limits are much
fainter than the loci of unreddened quasars with luminosities typical
of powerful radio sources at these redshifts. We have also been able
to detect the weakest quasars associated with such objects, like
3C~65, although at poorer limiting values of $A_V \simlt 5$. However,
the distribution of narrow emission line luminosities
(Fig.~\ref{fig:lines}) suggests that there are very few, if any, other
objects as weak as 3C~65 in the LRL sample at these redshifts.
Assuming then that the quasars within the many radio galaxies for
which we only have limits at $L'$ have luminosities typical of those
within other powerful radio sources, Fig.~\ref{fig:lz} makes it clear
that the extinction must be large in those sources we failed to detect
at $L'$.

Our data can be used to quantify the size of the dust-reddened quasar
population. Since all our sources are steep-spectrum objects, red
objects in our sample are unlikely to be due to beamed synchrotron
radiation, as has been suggested for the red flat-spectrum quasar
population (Serjeant \& Rawlings 1996; Benn et al.\ 1998). Among the
18 sources classified as quasars in Fig.~\ref{fig:lz}, there are two
(3C~190 and 3C~212; Smith \& Spinrad 1980) that fall into this
category. Scaling this to our complete sample of 12.5 `virtual'
quasars, and adding in 3C~22, 3C~41, and 3C~65, we find that
$28^{+25}_{-13}$\% of quasars are dust-reddened (90\% confidence
limits).

\subsection{Influence of quasar nuclei on the $K$--$z$ relation for 
radio galaxies}

Our first goal for this project was to investigate whether the
presence of reddened quasar light accounts for a significant fraction
of the total $K$-band flux in 3C radio galaxies. This could provide an
explanation for the brighter $K$ magnitudes displayed by 3C radio
galaxies compared to those from the fainter 6C and B2 samples (Eales
et al.\ 1997). We have therefore undertaken a similar analysis to that
of Eales et al., which involves fitting the least-squares regression
line to the 3C $K$-band data, and then examining the residuals about
this line for both the 3C and 6C samples (we do not supplement the
fainter sample with galaxies from the B2 survey). We use the same
63.9\,kpc metric aperture as Eales et al.\ and use the same
prescription for correcting measurements made in apertures of other
sizes.

We first confirm the representative nature of our sample by fitting a
regression line to the $K$-band magnitudes of our 18 galaxies. All 13
galaxies from the 6C survey in the same redshift range ($0.65 \leq z <
1.20$) lie above this line (i.e., they are fainter), with the mean
difference being 0.72\,mag.  This is in good agreement with the
0.59\,mag discrepancy found by Eales et al.\ (1997), and we can use a
Kolmogorov-Smirnov test to rule out the hypothesis that both samples
are drawn from the same distribution at better than 95\% confidence.

We then correct the photometry for the presence of reddened nuclear
light, using the fractional contamination determined in
Section~\ref{sec:hst} and repeat the process. This time the 6C
galaxies are on average only 0.56\,mag (40 per cent) fainter, although
the samples are still different at greater than 90\% confidence. Our
confidence levels are lower than those of Eales et al.\ (1997) solely
because our samples each contain only half the number of objects as
theirs.

\begin{figure}
\vspace*{130mm}
\includegraphics{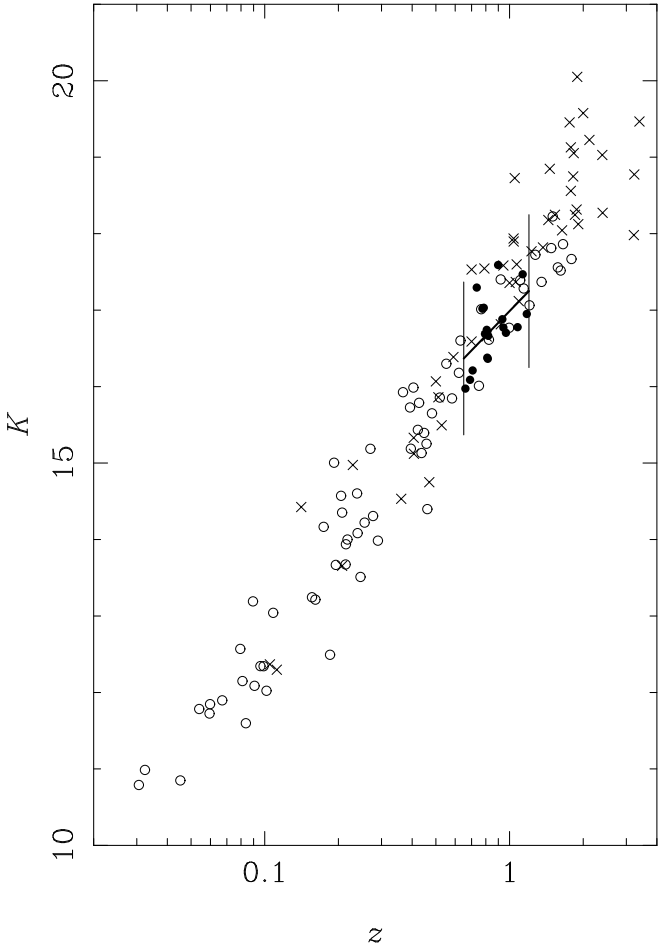}
\caption[]{Modified $K$-band Hubble diagram for radio galaxies from
the LRL and 6C samples (open circles and crosses respectively). The
solid circles are our sample of 18 galaxies with $K$-band photometry,
after correction for reddened quasar light. The two vertical lines
delineate the redshift range of our sample ($0.65 \leq z < 1.20$) and
the heavy line between them is the least-squares fit to the corrected
magnitudes. See Figure~\ref{fig:kz} for references for the
photometry.}
\label{fig:kzcorrect}
\end{figure}

Coupled with the conclusion of Leyshon \& Eales (1998) that scattered
light is not a major contributor to the total $K$-band flux of most 3C
radio galaxies, our analysis is able to rule out quasar light as the
cause of the correlation between radio luminosity and near-infrared
magnitude at $z \sim 1$. Emission lines can also be ruled out since
they would need to have a total flux of $\sim 2 \times
10^{-17}$\,W\,m$^{-2}$ to provide one-third of the total $K$-band
light --- this is several times greater than the flux of the
[O\III]~$\lambda\lambda$4959,5007 doublet yet the strongest lines
which fall into the $K$-band over our redshift range are intrinsically
much fainter (see also Rawlings et al.\ 1998a). Finally, the nebular
emission which has been suggested provides a significant fraction of
the optical light and might explain the alignment effect (Dickson et
al.\ 1995) is a negligible contributor at near-infrared
wavelengths. Assuming [O\II]/H$\beta \approx 4$ (McCarthy 1993), it
can contribute no more than about one per cent of the total $K$-band
flux.

The `extra' component in 3C radio galaxies, compared to 6C radio
galaxies, is therefore not directly related to the more luminous
central quasar, and must be additional stellar luminosity, and
presumably stellar mass, in the host galaxy. This fits together with
the smaller linear sizes of 6C radio galaxy hosts measured by Roche et
al.\ (1998). Plausible reasons why a correlation between host galaxy
mass and radio luminosity is seen at $z \sim 1$, but not at lower
redshifts, are given by Best et al.\ (1998).

Finally, we note that the r.m.s.\ deviation of the 3C galaxies about
the regression line is reduced from 0.59 to 0.43\,magnitudes when the
correction for non-stellar contamination is made. This reduced scatter
is the same as that observed around Eales et al.'s (1997)
least-squares regression line for $z < 0.6$.

\subsection{Test of the receding torus model}
\label{sec:rtorus}

In order to be able to compare the receding torus model to our
observational data, we have to associate lower limits to the nuclear
extinction with the radio galaxies which we failed to detect at
$L'$. Estimating these extinctions is a very uncertain affair, since
we do not know their intrinsic luminosities. Table~\ref{tab:intrinsic}
indicates that a `typical' source might have $L'_0 \approx 13$ ($K-L'
= 1.87$ for our canonical spectrum at all redshifts $z<1.2$), yet the
quasar within 3C~65 appears to be nearly ten times fainter, and the $z
= 0.768$ quasar 3C~175 is a magnitude brighter (Simpson,
unpublished). We adopt $L'_0 = 13.5$ so that the undetected sources
must have $A_{L'} \simgt 2$, or $A_V \simgt 15$ (at $z=1.2$; at lower
redshifts, the rest-frame visual extinction will be higher for the
same observed-frame $A_{L'}$) and use this limit for all 15 of our
non-detections.  Fig.~\ref{fig:lz} clearly shows the conservative
nature of this assumption.

\begin{figure}
\vspace*{123mm}
\includegraphics{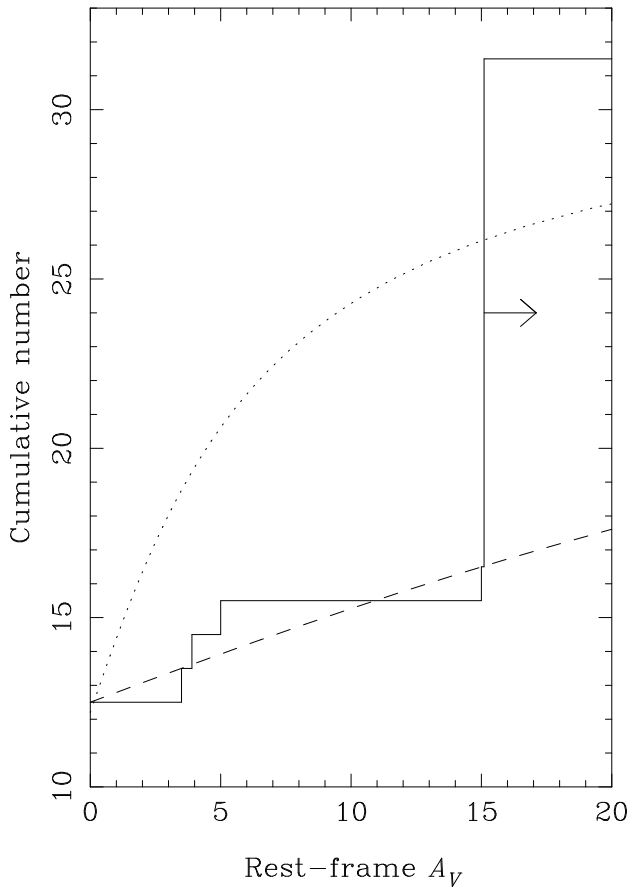}
\caption[]{Cumulative distribution of nuclear extinction for our
sample of radio sources (including the 12.5 `virtual' quasars). The
solid line shows the data and the dotted line shows the prediction of
the receding torus model of Hill et al.\ (1996). The dashed line shows
the prediction for the receding torus model, renormalized to the
statistics of our sample.}
\label{fig:torus}
\end{figure}

The mean radio luminosity of the radio sources with $0.65 \leq z <
1.20$ plotted in Fig.~\ref{fig:lines} is about 70 times greater than
that of Hill et al.'s (1996) $0.1 \leq z < 0.2$ sample. Hill et al.\
infer an opening angle of $20^\circ$ for their sample and, applying
the relationship $\tan \theta_{\rm c} \propto L_{178}^{0.3}$, we
expect $\theta_{\rm c} = 52^\circ$, in very good agreement with the
value derived directly from the quasar fraction in LRL. Using the same
value for the density of the torus, $n_{\rm H}$, as Hill et al., the
receding torus model predicts the distribution of extinctions shown by
the dotted line in Fig.~\ref{fig:torus}. Although the model does
predict fewer lightly-reddened sources than at low redshift ($\sim
30$\% with $A_V < 6$\,mag, as opposed to $\sim 50$\%), it still
overpredicts the number. By increasing the density of the torus by
about an order of magnitude above that derived for Hill et al.'s
sample, we can fit the data of Figure~\ref{fig:torus} with the dashed
line. Since this model is required to reproduce both the numbers of
naked and lightly-reddened quasars, there is very little additional
information in Fig.~\ref{fig:torus} with which to compare its
predictions. We cannot therefore claim any quantitative agreement
between the model and our observations.

There are several points to be made about the receding torus model.
First, one nust be careful when considering objects with low
extinction, since there may be additional obscuration which should not
be attributed to the torus. The line of sight to the quasar nucleus
could easily suffer a few magnitudes of visual extinction in its
passage through the host galaxy, or through foreground companion
objects such as those seen in our images of 3C~22. The one-sided radio
jet displayed by 3C~22 suggests a smaller angle between the radio (and
torus) axis and the line of sight than in 3C~41 and 3C~65, which have
similar nuclear extinctions, so it is quite possible that the torus
provides only a fraction (if any) of the obscuration. In fact, of the
four quasar nuclei we detect at $L'$, it is only 3C~265 that we can
say with reasonable certainty is seen through the heavy obscuration of
the torus.

The fact that we have needed to increase one of the parameters of the
receding torus model, namely the torus density $n_{\rm H}$, to allow
the model to fit our data, suggests that some refinements are needed.
The torus density could increase as a function of redshift, or quasar
luminosity (either directly or through the mass of the central black
hole or host galaxy), or both. Obviously this question can only be
answered by examining the distribution of extinctions in a sample
selected at a fainter radio flux limit, such as the 7C sample
(selected at a flux level 25 times fainter than LRL; Willott et al.\
1998a) although such a study is clearly beyond the capabilities of
UKIRT. We can, however, glean something just from the quasar fraction
in the 7C sample in the same redshift range, which is consistent with
the model's prediction of 11\% (Willott et al.\ 1998a). The relatively
small number statistics prevent this from being a definitive test,
although the agreement is certainly encouraging and suggests that the
torus height might be the same in all sources, irrespective of their
redshifts and luminosities.

\section{Utility of the thermal imaging technique}

The sensitivity of thermal imaging observations to point sources is
essentially limited by two factors. First, the size of the detection
aperture controls the detection threshold, and so the better the image
quality achievable, the greater the sensitivity. Secondly, there is
effectively a maximum integration time set by the stability of the
background from the telescope, and so the rate at which source photons
are detected is also important. Obviously then, the
diffraction-limited images that the new class of 8-m telescopes should
be able to obtain at $L'$ (${\rm FWHM} < 0\farcs2$) will enable much
deeper observations than we have been able to undertake.

The Infrared Camera and Spectrograph (IRCS; Tokunaga et al.\ 1998) to
be installed on the 8.3-m Subaru Telescope in 1999 will be able to
achieve a $3\sigma$ detection of a point source with $L' = 19.0$ in
1\,hour of integration, by using a 0\farcs5 detection aperture. For an
intrinsic magnitude of $L'_0 \approx 13$, this would allow the
detection of the quasar nucleus through as much as $A_V \approx
50$\,mag of obscuration. In reality, the sensitivity will be slightly
worse than this due to the presence of the host galaxy which is
expected to have $L' \approx 17$--18 within a 0\farcs5 aperture
(assuming $r_{\rm e} = 1\farcs5$ and $K - L' = 1$). However, the
excellent image quality should enable the nuclear point source, if
present, to be separated from the underlying galaxy with relative ease
and so much of the increased sensitivity will be retained.

The ability to measure obscuring columns to this depth would provide a
much more rigorous test of the receding torus model than could be
presented here. The model shown by the dashed line in
Fig.~\ref{fig:torus} predicts only six sources should be detected
with $15 < A_V < 50$, while the remaining nine will have even larger
obscurations. As was discussed in Section~\ref{sec:rtorus}, it is
important not to put too much weight on those sources with low
extinction ($A_V \simlt 5$) since some or even all of this material
may come from a source other than the torus. Given the apparent
evidence that no property of the torus is constant, the predictive
power of the receding torus model rests in its ability to infer the
obscuration towards the nuclei which are too heavily obscured to be
detected directly.

Improved sensitivity of the thermal-infrared imaging method will also
allow studies at much higher redshifts. This is an important extension
because there are obvious difficulties with forming enough dust for an
$A_V \sim 50$ torus in an arbitrarily short time, so there is likely
to be some critical redshift beyond which all radio sources have low
$A_{V}$.

Thermal-infrared imaging may also play an important r\^{o}le in
understanding the intrinsic orientation dependence of quasar
emissions. If the thermal-infrared flux arises from optically-thin
material and is isotropic, then statistical studies of, for example,
the $B - L'$ colours of $z \sim 1$ quasar nuclei could be used to test
whether the (rest-frame) UV continuum is also isotropic. Low-frequency
selected samples, such as those from the 3C and 7C samples, should
contain quasars randomly distributed in orientation with $\theta \leq
\theta_{\rm c}$, and would provide a suitable basis for such a
study. One possibility would be to look for a correlation between $B -
L'$ colour (corrected for reddening) and radio core dominance.
Various studies (e.g., Netzer 1987) have suggested that the UV
continua, and the low-ionisation broad emission lines arise in
optically-thick disks and thus emit anisotropically.

\section{Summary}

We have presented the results of a near- and thermal-infrared imaging
survey of a representative sample of 19 $z \sim 1$ radio galaxies from
the 3CR and 3CRR (LRL) samples. Four objects were detected at $L'$,
with red $K - L'$ colours which we attributed to partially-obscured
quasar light. We have used our infrared data to infer the nuclear
extinctions and contributions from non-stellar light at $K$ for these
four sources, and find the results to be in excellent agreement with a
more detailed analysis incorporating \HST\ data. We deduce that the
fraction of dust-reddened quasars at $z \sim 1$ is $28^{+25}_{-13}$\%
(90\% confidence). We find that the non-stellar light provides as much
as 80 per cent of the small-aperture $K$-band flux in two objects
classified as `radio galaxies', although it is unable to account for
the brighter $K$-magnitudes of 3C radio galaxies compared to 6C radio
galaxies at the same redshift, which must therefore be due to a
correlation between radio and host galaxy luminosities. We have also
investigated the observed distribution of nuclear extinctions in our
sample in the context of the receding torus model. Although the
smaller fraction of lightly-reddened nuclei is in qualitative
agreement with the predictions of the model, a quantitative match can
only by made by altering one of its parameters. We therefore conclude
that the receding torus model has limited predictive power when
comparing samples of radio sources at different redshifts and/or radio
luminosities, although we have been unable to determine whether it can
be applied to a single sample of sources with similar properties.
Deeper thermal-infrared imaging, which will be possible in the near
future, will be able to answer this and other questions.

\section*{Acknowledgments}

The United Kingdom Infrared Telescope is operated by the Joint
Astronomy Centre on behalf of the U. K. Particle Physics and Astronomy
Research Council. The authors are grateful to the UKIRT engineers and
especially to Tim Carroll, Tim Hawarden and Nick Rees for creating a
trouble-free night of observing with the fast guider on its first
night of public use. We also wish to thank Katherine Blundell and
Chris Willott for taking $L'$ data on 3C~22, Frossie Economou for
communicating the results on the spectra of 3C~41 and 3C~265 in
advance of publication, and Neil Medley, Sebastian Jester and Susan
Scott for checking the \HST\ model fitting. Both Chris Willott and the
anonymous referee are thanked for a critical reading of the manuscript
which led to substantial improvements. This work is based in part on
observations with the NASA/ESA {\it Hubble Space Telescope\/},
obtained from the data archive at the Space Telescope Science
Institute, which is operated by AURA, Inc.\ under NASA contract
NAS5--26555.

\end{document}